\documentclass{emulateapj}

\def\ligiant{\mbox{HKII~17435-00532}}

\def\kmsec{\mbox{km~s$^{\rm -1}$}}

\def\rpro{\mbox{$r$-process}}
\def\spro{\mbox{$s$-process}}
\def\ncap{\mbox{$n$-capture}}

\shorttitle{Lithium-, $s$-, $r$-Enhanced Metal-Poor Giant}
\shortauthors{Roederer et al.}

\begin{document}

\title{The Hobby-Eberly Telescope 
\textit{Chemical Abundances of Stars in the Halo} (\textit{CASH}) Project.\ 
\\ I.\ The Lithium-, $s$-, and $r$-Enhanced Metal-Poor Giant 
\ligiant\altaffilmark{1}}

\author{
Ian U.\ Roederer\altaffilmark{2},
Anna Frebel\altaffilmark{2,3},
Matthew D.\ Shetrone\altaffilmark{2,3},
Carlos Allende Prieto\altaffilmark{2},
Jaehyon Rhee\altaffilmark{4,5},
Roberto Gallino\altaffilmark{6,7},
Sara Bisterzo\altaffilmark{6},
Christopher Sneden\altaffilmark{2},
Timothy C.\ Beers\altaffilmark{8},
John J.\ Cowan\altaffilmark{9}
}

\altaffiltext{1}{Based on observations obtained with the Hobby-Eberly
Telescope, which is a joint project of the University of Texas at
Austin, the Pennsylvania State University, Stanford University,
Ludwig-Maximilians-Universit\"at M\"unchen, and
Georg-August-Universit\"at G\"ottingen.
}

\altaffiltext{2}{Department of Astronomy, University of Texas at Austin,
1 University Station, C1400, Austin, TX 78712-0259; 
iur,anna,shetrone,callende,chris@astro.as.utexas.edu}

\altaffiltext{3}{McDonald Observatory, University of Texas, Fort Davis, TX}

\altaffiltext{4}{Department of Physics, Purdue University, 
525 Northwestern Avenue, West Lafayette, IN 47907-2036; 
jrhee@physics.purdue.edu}

\altaffiltext{5}{Visiting Astronomer, Kitt Peak National Observatory, 
National Optical Astronomy Observatory, which is operated by the 
Association of Universities for Research in Astronomy, Inc.\ (AURA) 
under cooperative agreement with the National Science Foundation.}

\altaffiltext{6}{Dipartimento di Fisica Generale, Universit\`{a} 
di Torino, Torino, Italy; gallino,bisterzo@ph.unito.it}

\altaffiltext{7}{Centre for Stellar and Planetary Astrophysics, Monash 
University, PO BOX 28M, Clayton, VIC 3800 Australia}

\altaffiltext{8}{Department of Physics and Astronomy, 
Center for the Study of Cosmic Evolution, and 
Joint Institute for Nuclear Astrophysics, 
Michigan State University, East Lansing, MI; beers@pa.msu.edu}

\altaffiltext{9}{Homer L.\ Dodge Department of Physics and Astronomy, 
University of Oklahoma, Norman, OK 73019; cowan@nhn.ou.edu}

\begin{abstract}

We present the first detailed abundance analysis of the metal-poor
giant \ligiant.  This star was observed as part of the University of
Texas long-term project \textit{Chemical Abundances of Stars in the Halo}
(\textit{CASH}). A spectrum was obtained with the High
Resolution Spectrograph (HRS) on the Hobby-Eberly Telescope with a
resolving power of $R\sim15,000$.  Our analysis reveals that this
star may be located on the red giant branch, red horizontal branch, or
early asymptotic giant branch.  We find that this metal-poor
([Fe/H]~$=-2.2$) star has an unusually high lithium abundance
($\log\,\varepsilon\,($Li$)=\,+2.1$), mild carbon ([C/Fe]~$=+0.7$) and
sodium ([Na/Fe]~$=+0.6$) enhancement, as well as enhancement of both
\spro\ ([Ba/Fe]~$=+0.8$) and \rpro\ ([Eu/Fe]~$=+0.5$) material.  The
high Li abundance can be explained by self-enrichment through extra
mixing that connects the convective envelope with the outer
regions of the H-burning shell. If so, \ligiant\ is the most metal-poor 
star in which this short-lived phase of Li enrichment has been observed.  
The Na and \ncap\ enrichment can be explained by mass transfer from a 
companion that passed through the thermally-pulsing AGB 
phase of evolution with only a small initial enrichment of
\rpro\ material present in the birth cloud.
Despite the current non-detection of radial velocity variations 
(over $\sim180$ days), it is possible that \ligiant\ is in a long-period
or highly-inclined binary system, similar to other stars with similar 
\ncap\ enrichment patterns.

\end{abstract}


\section{Introduction}
\label{introduction}

The story of early Galactic nucleosynthesis is written in the chemical
compositions of very metal-poor halo stars.
The abundances in these stars reflect only a few chemical enrichment events, 
and hence this fossil record can be used to trace the chemical 
and dynamical evolution of the early Galaxy. 
While the individual abundances of most metals in these stars will remain 
unchanged throughout the stellar lifetimes, close examination is necessary 
to discern exceptions to this rule.

Lithium (Li), the only metal produced during the Big Bang, is often observed 
in unevolved metal-poor stars.  Information on the primordial Li abundance
can be inferred from, e.g., the Spite Plateau \citep{spite82}
Li abundance characteristic of most warm, metal-poor turn-off and 
subgiant stars.  
A wide spectrum of Spite Plateau/primordial Li abundances have been inferred 
from recent studies of these stars:
$\log\,\varepsilon\,($Li$) = 2.37$ (with an observational scatter
of 0.05-0.06~dex) \citep{melendez04},
$\log\,\varepsilon\,($Li$) = 2.21\,\pm\,0.09$ \citep{charbonnel05},
$\log\,\varepsilon\,($Li$) = 2.04$ or $2.15$ (depending on the range 
of metallicities of the stars included in the fit, since their 
Plateau has a metallicity dependence; \citealt{asplund06}), and
$\log\,\varepsilon\,($Li$) = 2.10\,\pm\,0.09$ \citep{bonifacio07}.\footnote{ 
We adopt the usual spectroscopic notations that
[A/B]~$\equiv$ log$_{\rm 10}$(N$_{\rm A}$/N$_{\rm B}$)$_{\star}$~--
log$_{\rm 10}$(N$_{\rm A}$/N$_{\rm B}$)$_{\sun}$, and
that log~$\varepsilon$(A)~$\equiv$
log$_{\rm 10}$(N$_{\rm A}$/N$_{\rm H}$)~$+$~12.00,
for elements A and B.}
Despite the wide range, the stellar result does not agree with the 
WMAP estimate of the primordial Li abundance,
$\log\,\varepsilon\,($Li$) = 2.64\,\pm\,0.03$ \citep{spergel07}.

Metal diffusion in long-lived, low-mass stars could present a 
solution to the problem for globular cluster stars
\citep{korn06}, while processing and depletion of $^{7}$Li by Population~III
stars prior to the formation of Population~II stars could explain 
the Li abundance found in present-day metal-poor field stars \citep{piau06}.
Some objects, however, appear to have
\textit{overabundances} of Li---in contrast to the sparse Galactic
production mechanisms of Li through cosmic ray spallation,
and in spite of the fact that Li is completely diluted and destroyed
in the stellar atmosphere by the time the star reaches the giant branch.

It was first proposed by \citet{cameron55} and \citet{cameron71} that
$^{7}$Li could be synthesized via the
$^{3}$He($\alpha$,$\gamma$)$^{7}$Be and
$^{7}$Be($e^{-}\nu$)$^{7}$Li reactions during the late stages of
stellar evolution, when a star is ascending the asymptotic giant branch
(AGB).  Here, the outer convection zone extends down into the
H-burning shell, which is enriched in $^{3}$He from proton-proton
chain reactions.
These nucleosynthesis reactions make it indeed possible for a star to
self-enrich its atmosphere with Li, but only for a very short period
of time before the freshly produced Li burns again. The enrichment
can be quite extreme, however, with Li abundances sometimes 
1--2~dex higher than the Spite Plateau value \citep[e.g.,][]{reddy05}.
In the past decade or so, a number of stars at roughly solar metallicity 
have been discovered that exhibit high Li abundances associated with 
self-enrichment.  These cases have been confirmed on theoretical grounds
\citep{charbonnel00} from their distinct position
on the Hertzsprung-Russell diagram.

At the heavy end of the periodic table, the neutron ($n$)-capture 
elements are also useful diagnostics of stellar interiors and nucleosynthesis.
During the last decade or so, new high resolution echelle spectrographs
on large aperture telescopes and new laboratory measurements of atomic 
data have revolutionized
the study of \ncap\ species in metal-poor stars
(see \citealt{sneden08} and references therein).
In most cases, slow ($s$)-process enrichment in metal-poor stars
is associated with binary systems, where the primary goes through 
the TP-AGB phase and produces large amounts of \spro\ material.
This material is transferred onto the lower-mass, longer-lived 
companion that is still observable.  
On the other hand, rapid ($r$)-process enrichment observed in metal-poor 
stars can be associated with explosive nucleosynthesis, although the 
exact site of the \rpro\ has not yet been conclusively identified.

Barium and europium are the heavy elements most commonly used to 
diagnose the \ncap\ enrichment history of a star because
they are so dominantly produced by the main 
$s$- and $r$-processes\footnote{
Two methods are commonly used to determine the relative 
contributions of the $s$- and $r$-processes to S.S.\ 
material, the classical method 
\citep{clayton61,seeger65,kappeler89} 
and the stellar model \citep{arlandini99}.
The classical method predicts that 85\% of the Ba in S.S.\ material
originated in the \spro\ and that 97\% of the Eu in S.S.\ material
originated in the \rpro\ \citep{simmerer04}.
The stellar model predicts that 81\% of the Ba in S.S.\ material
originated in the \spro\ and that 96\% of the Eu in S.S.\ material
originated in the \rpro.
}, respectively.
While [Ba/Fe] or [Eu/Fe] ratios can typically reveal the overall
content of $s$- and \rpro\ material in a star, 
the [Ba/Eu] ratio gives an indication of the relative contributions
of the $s$- and $r$-processes to the observed star.
Because of the physical conditions necessary to produce the \ncap\ 
species, it is probable that \textit{none} of the observed 
\ncap\ species in metal-poor stars were actually created by the
star they are presently observed in.
Some stars exhibit strong \spro\ enhancement, some exhibit strong
\rpro\ enhancement, and others exhibit significant amounts of 
both $s$- and \rpro\ material (see \citealt{jonsell06} and references
therein).

In this paper we present \ligiant, a metal-poor star that 
possesses an unexpectedly high (for its evolutionary state) Li abundance,
a significant amount of \spro\ material, and
a smaller but non-negligible amount of \rpro\ material.

\section{Observational Data and Measurements}
\label{data}

\subsection{Target Selection}
\label{targets}

The HK-II survey (\citealt{rhee01}; Rhee, Beers, \& Irwin, in prep.)
originated as an extension of the original HK objective-prism survey of
\citet{beers85,beers92}.  HK-II was designed to discover
a large sample of very metal-poor {\it red giant} stars with 
[Fe/H]~$\le -2.0$ by using {\it digitized} objective-prism spectra 
and 2MASS $JHK$ colors; many of these stars were likely to have been missed 
in the original (visual) selection of metal-poor candidates due to an 
unavoidable temperature bias.
The HK-II survey covers more than $\sim 7000$ deg$^2$ (one-sixth of the
entire sky), targeting the thick disk and halo of the Milky Way, over the
magnitude range $11.0 \le B \le 15.5$.  Ongoing
medium-resolution spectroscopic follow-up has newly confirmed 
more than 200 red giants and subgiants 
%some 130/130 red giants/subgiants 
with [Fe/H]~$\le -2.0$ in the first sample selected
from about 100 plates (Rhee \& Beers, in prep.).  

We have recently started the \textit{Chemical Abundances of Stars in the Halo}
(\textit{CASH}) Project with the 
Hobby-Eberly Telescope (HET; \citealt{ramsey98}) located at McDonald
Observatory.  This project aims to characterize the chemical composition of
the Galactic halo by means of abundance analyses of metal-poor
stars. Our goal is to build up, over the next several years, 
the largest high-resolution database available for these objects 
to investigate, 
\textit{inter alia}, the recent claim by \citet{carollo07} 
whether there is a chemical difference between the so-called 
``inner'' and ``outer'' halo populations.  
Furthermore, frequencies of stars with particular chemical abundances will be
established.
The very metal-poor giants identified in HK-II are one source of 
targets for our HET-\textit{CASH} Project,
and the spectrum of \ligiant\footnote{
In order to distinguish two source surveys, the stars from the HK-II
survey use the alphabetic prefix ``HKII'' while the stars from the original
HK survey use the prefixes ``BS'' and ``CS.''
The HK-II survey uses the exact same plates as the original HK survey, so
the first five digits (following the alphabetic prefix) in the star names
should be identical.  However, running ID numbers (last five digits) are
completely different.  For example, the star \ligiant\ is different
from the star BS~17435-532, but \ligiant\ \textit{is} 
a rediscovery of the star BS~17435-012, which was noted as being a 
metal-poor candidate during the initial visual inspection of the HK plates.
}
was taken as part of this project.  
The low metallicity of \ligiant\ was first identified in a
medium-resolution follow-up spectrum obtained in May 2005 at the 2.1\,m
telescope of Kitt Peak National Observatory.
Additional stars and larger samples will be presented in
separate papers.

\subsection{Observations and Data Reduction}
\label{observations}

The star \ligiant\ 
%(J2000.0 $11^{h}49^{m}03.3^{s} +16^{\circ}58'41.7''$)
was observed on 2007 February 09 and 28 with the
high resolution spectrograph (HRS; \citealt{tull98}) 
at the HET at McDonald Observatory as part of normal 
queue mode scheduled observing \citep{shetrone07}.
Our spectra have $R \sim 15,000$ and were taken through the 
3'' slit with the 316g cross-disperser setting. 
The spectral coverage is 4120 to 7850\,{\AA}, with a small break
from 5930 to 6030\,{\AA} resulting from the gap between the blue and
red CCDs of the HRS. 
One 10\,min exposure was taken on each night; 
we scheduled the two visits at least two weeks apart 
to test for radial velocity variations. 

For the reduction of all HET/HRS data obtained through the \textit{CASH}
Project we set up the IDL-based REDUCE data reduction software package
\citep{piskunov02}. 
The extracted spectra were combined after correcting for any
heliocentric radial velocities.
Overlapping echelle orders were merged together to produce the final
spectrum. 
The final $S/N$ values per pixel measured from clean,
line-free regions of the continuum in the REDUCE spectrum
range from $\sim50/1$ at 4480\,{\AA} to $\sim130/1$ at 5730\,{\AA} to
$\sim160/1$ at 6700\,{\AA}.  
More details on this reduction procedure for the entire 
HET-\textit{CASH} sample 
will be given in Frebel et al.\ 2008 (in prep.).

We also reduced
the data with standard IRAF\footnote{IRAF is distributed by the
National Optical Astronomy Observatories, which are operated by the
Association of Universities for Research in Astronomy, Inc., under
cooperative agreement with the National Science Foundation.} routines
in the \textit{echelle} and \textit{onedspec} packages for overscan
removal, bias subtraction, flat fielding, scattered light removal, and
order extraction.
Individual orders were not merged together.  
The final $S/N$ values from the IRAF spectrum are very 
comparable to the REDUCE spectrum.
Equivalent widths were also measured for 121 lines in common to the
two spectra to test any systematic differences.  
We find a mean offset of $\Delta=-3\pm8$\,m{\AA} 
(where $\Delta$ is defined as EW$_{\rm REDUCE}-$EW$_{\rm IRAF}$). 
There are no statistically significant trends with either
equivalent width or wavelength between the two differently-reduced
sets of spectra.

To obtain additional information on the radial velocity of
\ligiant, two supplemental exposures were taken on 2007 July 15 and
2007 August 05 with the Cross-Dispersed Echelle Spectrometer \citep{tull95}
on the Harlan~J.~Smith 2.7\,m Telescope at $R\,\sim\,60,000$.  
Reduction was performed using the standard IRAF routines.
The exposure times were 1200 seconds, yielding $S/N \sim 10/1$
at 4480\,{\AA}, $\sim 20/1$ at 5730\,{\AA}, and $\sim 25/1$ at 6700\,{\AA}.
Since the $S/N$ of the 2.7\,m spectra is much less than the $S/N$ 
of the HET/HRS spectra,
these spectra were solely used for the radial velocity analysis
and not for the abundance analysis.

\subsection{Radial Velocity}
\label{rv}

The radial velocities were measured in the IRAF reduced spectrum by
combining the orders covering the wavelengths 4900 to 5800\,{\AA},
which are mostly clear of telluric features and sky emission lines.
The spectra were then cross-correlated against the Arcturus atlas
\citep{hinkle00}
spectra using the IRAF task \textit{fxcor} to yield a relative velocity.
Heliocentric corrections were made using the IRAF task
\textit{rvcorrect}.  We also cross-correlated the telluric A band
against a model of this band we constructed to yield a zero point
correction to the wavelength scale; this is necessary because the
ThAr fibers are separate from the science fibers in the HRS.
In Table~\ref{rvdata} we have
listed the epochs of observations as well as the resulting velocities.
The weighted average velocity is 38.9\,km\,s$^{-1}$ with an error of
the mean of 0.3\,km\,s$^{-1}$.
The errors in the individual velocity measurements include the
systematic errors from the telluric features and the error in the
relative velocity with respect to Arcturus.
All of the velocity measurements are within
1\,$\sigma$ of the average velocity.  The first and last observations
of this star are spaced nearly six months apart and successive
observations are separated by more than two weeks.  We find no
evidence from these radial velocity measurements that \ligiant\ is in
a binary system.  If it were in a binary, it must have a very long period 
and/or an amplitude smaller than $\sim$~1.0\,km\,s$^{-1}$, a
result of a face-on orbit with very large sin\,$i$.

\subsection{Atmospheric Parameters}
\label{atmosphere}

Basic stellar data for \ligiant\ are shown in Table~\ref{basicdata}.
For our analysis, we use the most recent version (2002) of the LTE
spectrum analysis code MOOG \citep{sneden73}. 
We use model atmospheres computed from the
\citet{kurucz93} grid without convective overshooting.
Interpolation software for the Kurucz grid has been
kindly provided by A.\ McWilliam and I.\ Ivans (2003,
private communication).
We measure equivalent widths of 
86 Fe~\textsc{i} lines and 6 Fe~\textsc{ii} lines
in \ligiant\ by fitting Gaussian profiles
with the SPECTRE code \citep{fitzpatrick87}.
Our equivalent width measurements for all species in 
\ligiant\ are presented in Table~\ref{ewmeasure}.

The \citet{schlegel98} dust maps predict $E(B-V)=0.042$ in the 
direction of \ligiant.
This is the expected reddening value at infinite distance.
At the high Galactic latitude of our halo star, $b = +73^{\circ}$, 
its distance implies that the reddening along the line-of-sight
to this star is essentially the reddening at infinity; cf.\
Equation~4 of \citet{mendez98}.
The earlier predictions of \citet{burstein82} yield $E(B-V) \lesssim 0.01$.
We also attempt to use the \citet{munari97} calibrations between the
interstellar Na~D1 line equivalent width and reddening.  
The interstellar Na~D lines suffer from telluric emission and 
blending with the stellar lines in our HET spectra (and no simultaneous
sky spectra were obtained), and these lines fall between echelle orders 
in our higher-resolution McDonald spectra.
Given these difficulties, our interstellar Na~D1 equivalent width
should be interpreted as an upper limit.  
If we naively adopt this equivalent width, the \citet{munari97}
calibrations then predict an uninteresting upper limit for the
reddening of $E(B-V) \lesssim 0.12$, and larger equivalent width 
estimates would increase the reddening upper limit.
Lacking further information, we adopt the \citet{schlegel98} 
reddening estimate.

We derive the effective temperature of \ligiant\ 
from recent color-$T_{\rm eff}$ calibrations of \citet{alonso96,alonso99}.
We collected $V$, $I$ (TASS: The Amateur Sky Survey, version 2; 
\citealt{droege06}), $J$, $H$, and $K$ (2MASS\footnote{
This research has made use of the NASA/IPAC Infrared 
Science Archive, which is operated by the Jet Propulsion 
Laboratory, California Institute of Technology, under 
contract with the National Aeronautics and Space Administration.}
\footnote{
This publication makes use of data products from the Two Micron 
All Sky Survey, which is a joint project of the University of 
Massachusetts and the Infrared Processing and Analysis 
Center/California Institute of Technology, funded by the 
National Aeronautics and Space Administration and the 
National Science Foundation.}; \citealt{skrutskie06})
broadband photometry measurements from the literature.\footnote{\label{bvri}
New $BVRI$ broadband photometry of \ligiant\ was obtained by 
T.~Chonis and M.~Gaskell after our analysis was completed.
The $(V-I)$ color computed from this work, 0.902, is in superb
agreement with the $(V-I)$ color computed from the TASS photometry, 0.908. 
This agreement reinforces our confidence in the reliability of the
colors employed for our photometric temperature determination,
especially in light of the rather large uncertainty in the 
$V$ magnitude from TASS, $V=13.145\,\pm\,0.166$.}
Using the dereddened $(V-I)$, $(J-H)$, $(J-K)$, and $(V-K)$ colors 
derived from TASS and 2MASS photometry, we derive a mean temperature 
of 5173\,K with a standard deviation of 82\,K.\footnote{
If we include the $(B-V)$, $(V-R)$, and $(R-I)$ colors from the new
$BVRI$ photometry described in footnote~\ref{bvri} in place of 
TASS photometry, the mean of the seven calibrations rises to 5200\,K.
Despite the dangers of mixing sources of $BVRI$ photometry, if
we adopt $B$ and $R$ from the new photometry and $V$ and $I$ from
TASS, the mean only changes slightly to 5228\,K.}
To estimate the amount of systematic uncertainty, we compare with
the temperature scale defined by the \citet{ramirez05a,ramirez05b}
$(V-J)$, $(V-H)$, and $(V-K)$ calibrations; at the metallicity
of \ligiant, our $(V-I)$ falls beyond the range of applicability
of their calibrations.
These three color-$T_{\rm eff}$ calibrations predict a mean of 5079\,K,
about 100\,K lower than the \citet{alonso99} mean.
If we were to adopt zero reddening or twice the reddening predicted by the
\citet{schlegel98} maps, our temperatures would differ by roughly 130\,K.
Considering all of these matters, we round the \citet{alonso99}
predicted temperature to 5200\,K and adopt a conservative total 
uncertainty of 150\,K.

%We use the infrared flux method color-$T_{\rm eff}$ calibration  
%of \citet{alonso96,alonso99} and the recalibration by
%\citet{ramirez05a,ramirez05b}.
%The temperatures derived from these colors are presented in
%Table~\ref{colorteff}.
%For $(V-K)$, the predicted temperatures are in good agreement
%between the two systems.
%The temperatures derived from the three \citet{alonso99} 
%color calibrations---two of which do \textit{not} involve the $V$ 
%magnitude---are also in good agreement, suggesting that the $V$ magnitude for 
%\ligiant\ is reliable for our purposes.
%We adopt the \citet{alonso99} $(V-K)$ temperature,
%5200\,K, which lies in the middle of the range of all of these values.

Our method for deriving the surface gravity requires that, 
for our adopted effective temperature, the 
abundances derived from neutral and ionized lines of Fe agree.
We neglect the effects of non-LTE line formation, which 
may affect the Fe~\textsc{i} abundance measurements at the 
$\sim0.1$~dex level \citep[e.g.,][]{asplund05a}.
The microturbulence is measured by requiring that 
the abundances derived from strong and weak lines of Fe~\textsc{i} agree.
These parameters are varied iteratively 
until we arrive at our final values.
[Fe/H] is defined by the Fe abundance
for the final set of $T_{\rm eff}$, $\log(g)$, and $v_{t}$.
We increase the overall metallicity $Z$ of the atmosphere
by $+0.4$~dex to account for the extra electrons from the
$\alpha$-elements that contribute to the H$^{-}$ continuous opacity;
this artificial increase alters our derived abundances of the
electron-donating elements by $\lesssim 0.01$~dex and is 
completely negligible when elemental abundance ratios are considered
(see, e.g., \S~4.3 of \citealt{roederer08}).
We adopt the following atmospheric parameters for \ligiant:
$T_{\rm eff}/\log(g)/v_{t}/$[Fe/H]~$=
5200$\,K$/2.15/2.0\,$km\,s$^{-1}/-$2.23.
These values are also displayed in Table~\ref{basicdata}.

An alternate method for deriving $T_{\rm eff}$, $\log(g)$, and 
[Fe/H] uses the HET calibration sample for the Sloan Digital Sky Survey
to fit a region around the Mg~~\textsc{i}~b triplet
(see discussion in \citealt{allendeprieto08}).
For the sum of the two individual spectra of \ligiant, this method
finds $T_{\rm eff}/\log(g)/$[Fe/H]~$=
4978\,\pm\,72$\,K\,$/2.06\,\pm\,0.13/-2.26\,\pm\,0.06$.  
While the gravity and metallicity are in good agreement with our 
spectroscopic values, the temperature is about 200\,K cooler.
%Given the range of photometric $T_{\rm eff}$ values and this 
%value, we are inclined to adopt an uncertainty of $\pm\,150$\,K 
%for our ($V-K$) temperature.
The gravity obtained from the Fe ionization balance is also
not well-constrained because we are only able to measure equivalent widths
of 6 Fe~\textsc{ii} lines, and we adopt an uncertainty of $\pm\,0.4$ 
in $\log(g)$.
We also adopt $\pm\,0.3$\,km\,s$^{-1}$ as the uncertainty in $v_{t}$.
The uncertainty in the metallicity, $\pm\,0.23$~dex, is found by combining
the uncertainties in the other atmospheric parameters and the 
line-to-line Fe~\textsc{i} abundance scatter in quadrature.

\section{Evolutionary Status of \ligiant}
\label{hrdtext}

\begin{figure}
%\epsscale{0.77}
\epsscale{1.00}
%\plotone{f1.bw.eps}   %% b/w figure, print edition
\plotone{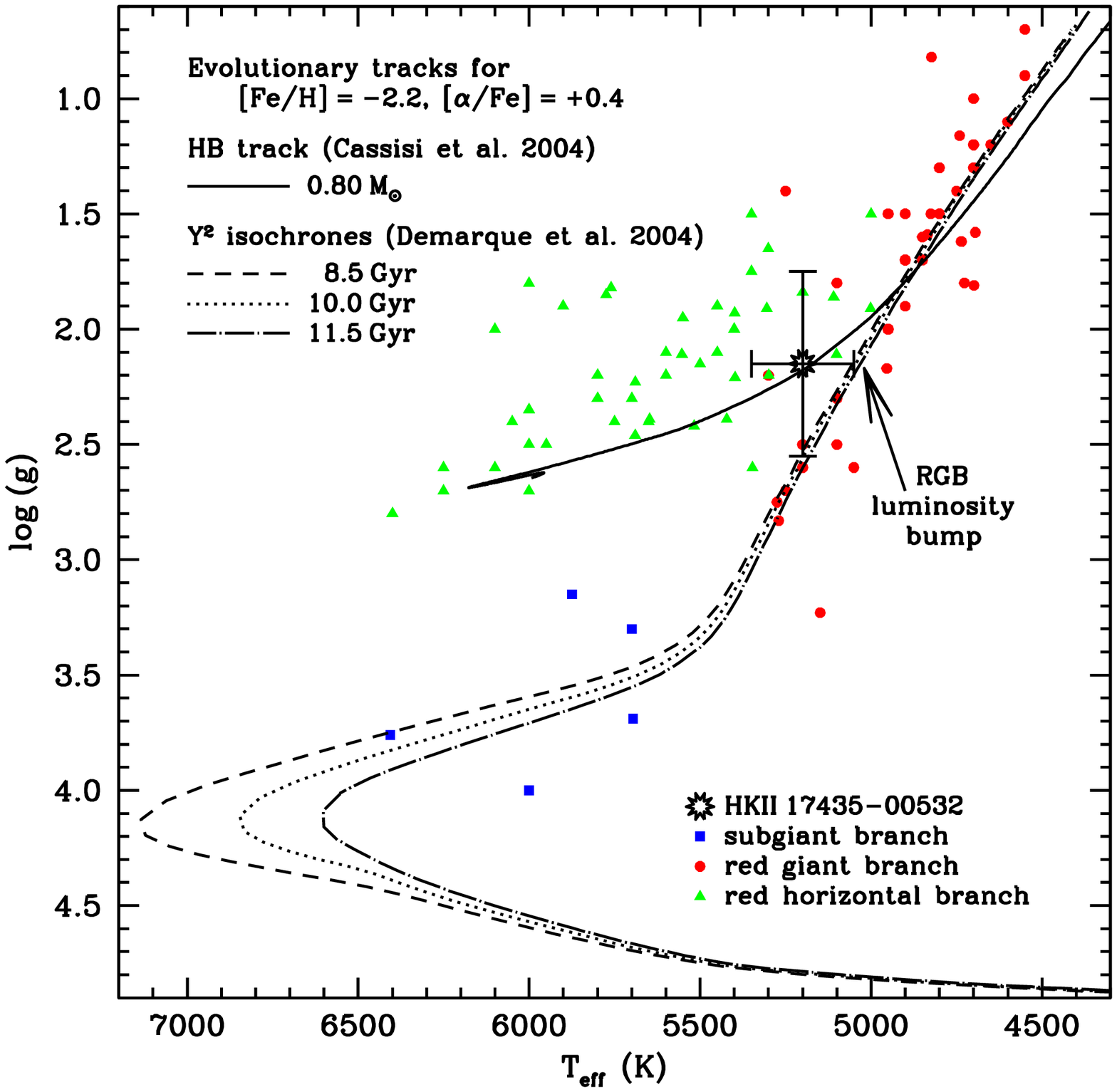}   %% color figure, online edition only
\caption{
\label{hrd}
Spectroscopic gravities are shown as a function of effective temperature
for \ligiant\ and a sample of other evolved metal-poor stars 
from previous studies.
\ligiant\ is indicated by the sunburst.
The evolved sample is compiled from the data in 
\citet{behr03} (only stars with [Fe/H]~$<-1.0$),
\citet{cayrel04}, and \citet{preston06}.
Filled squares indicate stars classified as ``turn-off'' or 
``subgiant'' stars (blue in the online edition),
filled circles indicate stars classified as being on the RGB
(red in the online edition), and 
filled triangles indicate stars classified as being on the RHB
(green in the online edition).
We also display several evolutionary tracks for reference,
which have all been computed for [Fe/H]~$=-2.2$ and 
[$\alpha$/Fe]~$=+0.4$.
Three sets of $Y^{2}$ isochrones \citep{demarque04} are 
displayed, corresponding to ages of 8.5\,Gyr (dashed line),
10.0\,Gyr (dotted line), and 11.5\,Gyr (dash-dotted line).
The assumed age has little effect on the location of the RGB.
A synthetic HB \citep{cassisi04} for 
$M=0.80\,M_{\sun}$ is shown by the solid line.
Small changes in the assumed mass have little effect 
on the location of the HB.
The arrow indicates the location of the RGB luminosity bump.
[See electronic edition for a color version of this figure.]
}
\end{figure}

The location of \ligiant\ on the $T_{\rm eff}$-$\log(g)$ diagram 
is shown in Figure~\ref{hrd}.
We plot the spectroscopically-determined gravity %of $\log g = 2.15$ 
as a function of effective temperature for \ligiant\ 
and many other evolved stars collected from previous studies.
\ligiant\ is located in the region of the diagram that is
populated by metal-poor stars that have been classified
as being on the red giant branch (RGB) and stars that have been classified
as being on the red horizontal branch (RHB). 
For reference, we also display evolutionary tracks in Figure~\ref{hrd}.
The $Y^{2}$ isochrones \citep{demarque04}
for three different ages (8.5, 10.0, and 11.5\,Gyr) with
metallicity [Fe/H]~$=-2.2$ and [$\alpha$/Fe]~$=+0.4$,
similar to that of our star, are shown. 
As can be seen, these isochrones approximately match the positions of the
stars on the subgiant and red giant branches.
For an old, metal-poor population of giants, 
it is clear that 
the assumed age has no effect on their location in the 
$T_{\rm eff}$-$\log(g)$ diagram.

We also show a synthetic horizontal branch (HB) track \citep{cassisi04}
for $M=0.80\,M_{\sun}$.
At low metallicity, these tracks are only available for
$Y=0.23$, [$\alpha$/Fe]~$=0.0$, and limited values of the metallicity $Z$.
To scale $Z$ of these tracks for an $\alpha$-enhancement
of $+0.4$, we use the formula given in \citet{kim02}
\citep[see also, e.g.,][]{salaris93}.
A metallicity [Fe/H]~$\approx-2.2$ with [$\alpha$/Fe]~$=+0.4$
corresponds to $Z\approx 2.6 \times 10^{-4}$, which
we obtain from the \citet{cassisi04} tracks
by interpolation between the $Z= 1 \times 10^{-4}$ and
$Z= 3 \times 10^{-4}$ tracks.
We follow the prescription given in \citet{preston06} to convert
the $\log(L/L_{\sun})$ given by \citet{cassisi04} to
$\log(g/g_{\sun})$.
On the scale of this figure,
the location of the HB is very insensitive
to small changes in the assumed mass for stars on the HB.
\citet{preston06} performed a comparison of several different
sets of HB tracks to the \citet{cassisi04} set.
They found differences in $\log(L)$ (which is equivalent to $-\log(g)$) 
to be $\lesssim0.15$~dex, which is well within the uncertainty 
in our $\log(L)$ determination for \ligiant.

%%%%%%%%%%%%%%%%%%%%%%  SAVE FOR REFERENCE!!!  %%%%%%%%%%%%%%%%%%%%%%%
%% For [Fe/H]=-2.2, [a/Fe]=+0.4 ==> Z_(alpha-corrected) = 0.00026.  %%
%% The HB tracks are only available with Z = 0.0001 and 0.0003.     %%
%% Interpolate to find 0.0026 values.                               %%
%%%%%%%%%%%%%%%%%%%%%%%%%%%%%%%%%%%%%%%%%%%%%%%%%%%%%%%%%%%%%%%%%%%%%%

\ligiant\ coincidentally appears to lie on the $M=0.80\,M_{\sun}$ HB track;
however, due to the large uncertainties in our determination of the
temperature and surface gravity of this star, 
we can not draw any firm conclusions from this.
This star could possibly be
ascending the RGB for the first time
or ascending the early AGB from the HB.
A higher gravity or cooler temperature would place the star on the RGB.  
A lower gravity or warmer temperature would place the star above the 
$M=0.80\,M_{\sun}$ HB track in the region of the diagram
populated by (presumably) lower-mass RHB stars
(cf.\ Figure~15 of \citealt{preston06}).
This is reasonable because other metal-poor stars are 
found in this region. 
\ligiant\ cannot be in the thermally-pulsing (TP)-AGB phase, 
which would require
$\log(g) \lesssim 0.9$ and $T_{\rm eff} \lesssim 4800$\,K
(see, e.g., Figure~3 of \citealt{masseron06}).
To achieve Fe ionization balance at $\log(g)=0.9$
would require a non-LTE correction of $\sim-0.5$~dex to the
Fe~\textsc{i} abundance, which is much greater than 
has been suggested by \citet{asplund05a}.

From fundamental relations, we calculate
$\log(L/L_{\sun})=2.01\,\pm\,0.50$ for \ligiant, assuming
%$T_{\rm eff}\,=\,5200\,\pm\,150$\,K,
%$\log(g)\,=\,2.15\,\pm\,0.4$, and
$M\,=\,0.8\,\pm\,0.1\,M_{\sun}$ for this star and
$T_{\rm eff}\,=\,5780$\,K and
$\log(g)\,=\,4.44$ for the Sun.
The uncertainty in $\log(L)$ is dominated by our uncertainty in the gravity.

\section{Validation of Our Abundance Analysis Techniques}
\label{validation}

We chose suitable elemental lines for our abundance analysis in the
range of $\sim4120$\,{\AA} to $\sim7850$\,{\AA} from the extensive
linelists of six recent studies of abundances in very metal-poor stars
\citep{fulbright00,cayrel04,honda04a,barklem05,ivans06,frebel07}.  
We adopt the log($gf$) values employed by these studies for all species 
except Cr~\textsc{i}, whose log($gf$) values were recently redetermined
by \citet{sobeck07}.  
To confirm the integrity of our linelist 
and validate our abundance analysis method for
measuring chemical compositions, we carried out a basic
abundance analysis of the well-studied cool giant HD~122563 and the
warm main-sequence turn-off star HD~84937.  The spectra have
$R\,\sim\,80,000$, very high $S/N$, and were taken from the VLT-UVES
archive.

We measure equivalent widths for 227 Fe~\textsc{i} and
35 Fe~\textsc{ii} lines in HD~122563 and
197 Fe~\textsc{i} and 26 Fe~\textsc{ii} lines in HD~84937.
We derive $T_{\rm eff}/\log(g)/v_{t}/$[Fe/H]
$=4570\,\pm\,100$\,K$/0.85\,\pm\,0.3/2.0\,\pm\,0.3$\,km\,s$^{-1}/-2.81\,\pm\,0.15$
for HD~122563
and $6300\,\pm\,100$\,K$/4.0\,\pm\,0.3/1.2\,\pm\,0.3$\,km\,s$^{-1}/-2.28\,\pm\,0.12$
for HD~84937 using the methods described in \S~\ref{atmosphere}.
In Table~\ref{atmcompare} we compare our derived parameters with
a number of other recent high-resolution studies.
For both HD~122563 and HD~84937, each of our derived parameters agree 
with the mean from other studies within their mutual 1$\sigma$ 
uncertainties.
%(see, e.g., \citealt{johnson02a} and \citealt{honda04a,honda04b} for
%recent high-resolution analyses of HD~122563 and
%\citealt{smith93}, \citealt{fulbright00},
%and \citealt{gratton03} for HD~84937).

\begin{figure}
\epsscale{1.00}
\plotone{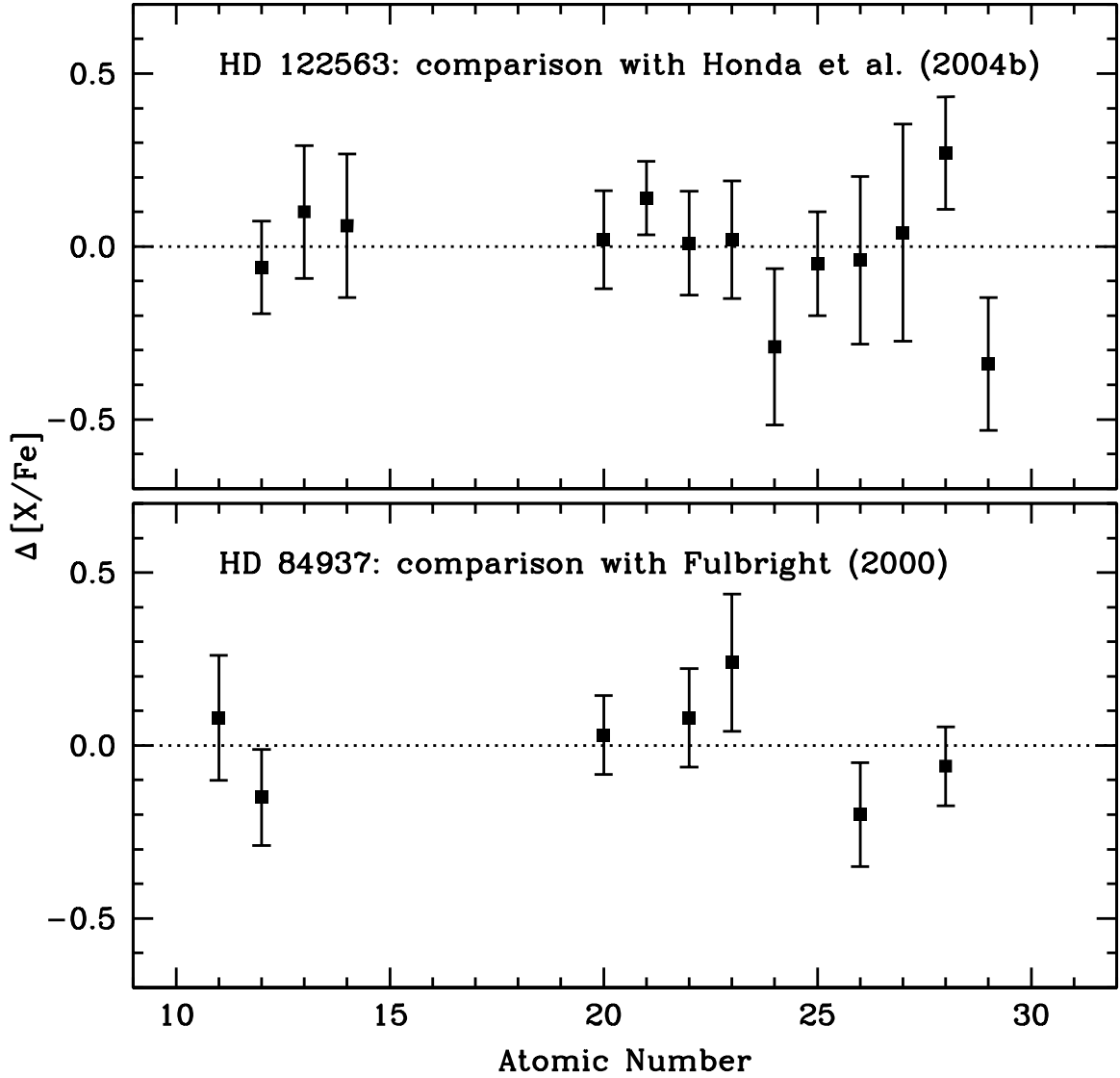}
\caption{
\label{comparison}
Comparison of the abundances derived in our study and two previous
studies of HD~122563 and HD~84937, \citet{honda04a,honda04b} 
and \citet{fulbright00}.
$\Delta$\,[X/Fe] is in the sense of (our study)$-$(other study).
For Fe, we show $\Delta$\,[Fe/H] rather than $\Delta$\,[X/Fe].
}
\end{figure}

We also measure 204 equivalent widths for 17 other species
($11 \leq Z \leq 30$)
in HD~122563 and 161 equivalent widths for 16 other species in HD~84937.
In Figure~\ref{comparison} we compare our derived abundances
in these two stars with the abundances derived by two previous studies.

In HD~122563, the derived abundances agree with the
\citet{honda04a,honda04b}
abundances within the uncertainties for all species except
Cr, Ni, and Cu, which differ by $-$0.29, $+$0.27,
and $-$0.44~dex, respectively, where the differences are in the sense
of (our study)$-$(other study).
We measured equivalent widths for 16 Cr~\textsc{i} lines and
5 Cr~\textsc{ii} lines, whereas \citet{honda04a,honda04b} only
used 3 and 2 lines of these species, respectively.
For Cr, only 0.02~dex of the difference can be accounted for
by our use of updated Cr~\textsc{i}
$\log(gf)$ values from \citet{sobeck07}.
Another 0.14~dex results from the different atmospheric parameters
derived in these two studies.
This is sufficient to bring the two Cr abundance measurements
into agreement within the uncertainties. 
\citet{honda04a,honda04b} derive the Ni abundance from two strong lines,
whereas we measure 21 Ni lines. 
The equivalent widths of the two lines are common to both studies 
are in good agreement, therefore we attribute the discrepancy in
the derived Ni abundances to the number of lines measured.
The Cu abundance in both the \citet{honda04a,honda04b} studies
and our study was measured from the
Cu~\textsc{i} line at 5105\AA; the difference in the
derived abundances can be traced to the different equivalent
width measurements of this very weak line, 3.3 and 1.7\,m\AA, respectively.

In HD~84937, the derived abundances agree with the \citet{fulbright00}
abundances within the uncertainties for all species except
Mg, V, and Fe, which differ by $-$0.15, $+$0.24, and
$-$0.20~dex, respectively, again in the sense of (our study)$-$(other study).
We selected the \citet{fulbright00} study for comparison because---in
our efforts to compare with a set of homogeneous measurements---it 
offered the most elements in common with our measurements.
The Mg and V abundance discrepancies can be explained by the different
atmospheric parameters, which account for 0.04~dex and 0.06~dex 
of the difference, respectively, 
reducing the discrepancies to less than the uncertainties.
When compared with many other studies (in Table~\ref{atmcompare}),
our [Fe/H] is in agreement with these studies within their mutual
1$\sigma$ uncertainties, so we do not consider it to be discrepant.
%The [Fe/H] derived by \citet{fulbright00} appears to be
%systematically higher than the [Fe/H] value for HD~84937
%found by many other recent studies
%(\citealt{fulbright00}: [Fe/H]~$=-2.08$;
%\citealt{smith93}: [Fe/H]~$=-2.4$;
%\citealt{gratton03}: [Fe/H]~$=-2.22$;
%this study: [Fe/H]~$=-2.28$).
%\citet{fulbright00} adopted a slightly higher temperature (6375\,K)
%than we did (6300\,K).
%If we use the atmospheric parameters derived by \citet{fulbright00},
%we find a higher Fe abundance
%([Fe/H]~$=-2.12$) than if we use our atmospheric parameters,
%which accounts for the discrepancy.
%This systematic difference disappears when [X/Fe] abundance ratios
%are considered for the other species.

We also smooth the UVES spectra of HD~122563 and HD~84937
to $R\,\sim\,15,000$ to simulate the spectral resolving power
employed in our study.
We do not degrade the $S/N$ of these spectra to match the $S/N$ of 
our HET spectra; the purpose of smoothing the spectra is to 
allow us to identify lines that are blended at the lower
spectral resolution and remove them from our final linelist.
We accomplish this by noting the spuriously-high abundances
produced by the blended lines in a line-by-line comparison with 
the abundances derived from the higher resolution spectra.
We then measure equivalent widths from unblended lines in each star
and re-derive elemental abundances using the same atmospheric parameters. 
The abundance ratios for both stars are identical to the
abundance ratios determined from the $R\,\sim\,80,000$ spectra
within the measurement uncertainties. 
From this and other comparisons given above, we conclude that our
linelist and abundance analysis technique are reliable.

\section{Abundance Analysis}
\label{abundance}

\subsection{Comments on Individual Species}
\label{comments}

In Table~\ref{abund} we list our derived LTE abundances, along with
non-LTE corrections when available, for 23 elements in \ligiant.  
We reference the elemental abundance ratios to the solar photospheric 
abundances given by \citet{grevesse02}.  These abundances have been 
derived primarily from analyses using 1D hydrostatic model atmospheres
under the assumption of LTE, as we have done in our current study of
\ligiant.  

\begin{figure}
%\epsscale{0.62}
\epsscale{1.00}
%\plotone{f3.bw.eps}   %% b/w figure, print edition
\plotone{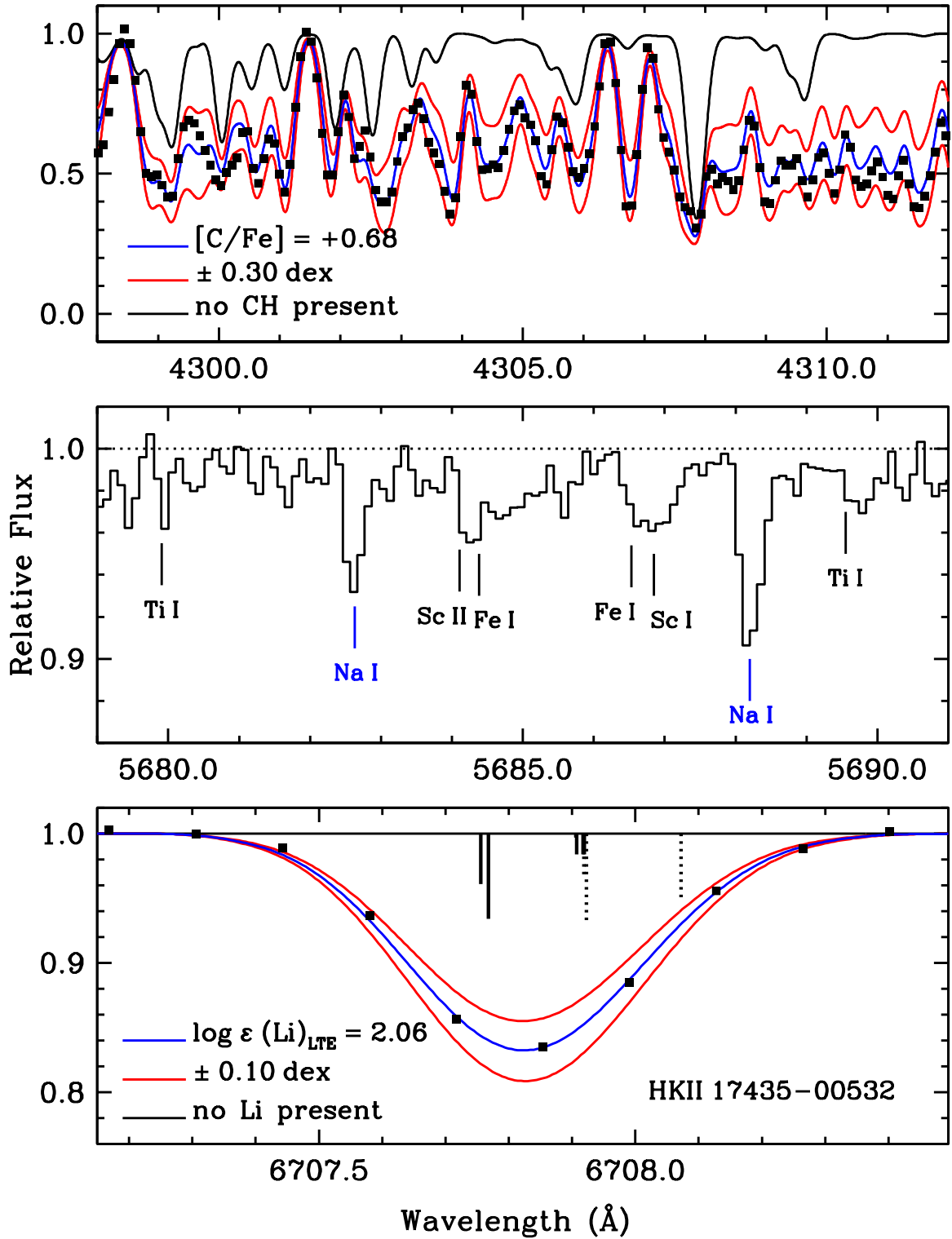}   %% color figure, online edition only
\caption{
\label{specplot}
Portions of our spectrum of \ligiant.
The top panel displays part of our synthesis of the CH G-band
near 4300\,{\AA},
the middle panel shows the spectral window around the 
Na~\textsc{i} 5682 and 5688\,{\AA} lines, and 
the bottom panel shows our synthesis of the Li~\textsc{i} 6707\,{\AA}
resonance line.
In the top and bottom panels, the solid line (blue in the 
online edition) represents our best fit synthesis, while the 
dashed lines (red in the online edition) represent variations in the
best-fit abundance.
The dot-dashed line (black in the online edition) 
represents a synthesis with no CH or Li present.
The observed spectrum is indicated by solid squares.
In the middle panel, the observed spectrum is indicated by the histogram
and the dotted line indicates the location of the continuum.
In the bottom panel, the relative strengths and positions of the 
hyperfine components of $^{7}$Li are represented by solid sticks,
while the $^{6}$Li components are represented by dotted sticks. 
We can assume that only $^{7}$Li is present without altering 
any of our conclusions.
[See electronic edition for a color version of this figure.]
}
\end{figure}

We synthesize the Li~\textsc{i} 6707\,{\AA} doublet using the linelist
of \citet{hobbs99}, who compiled the 
hyperfine structure component wavenumbers of this resonance line
from measurements by \citet{sansonetti95} and the
oscillator strength measurements of \citet{yan98}.
For the purposes of this paper, we can assume that all of the 
Li present is $^{7}$Li without altering any of our conclusions.
Our synthesis of this line is shown in Figure~\ref{specplot}.
We derive $\log\,\varepsilon\,($Li$)_{\rm LTE}\,=\,2.06$.
Using the \citet{carlsson94} non-LTE corrections for 
the 6707\,{\AA} line, we find 
$\log\,\varepsilon\,($Li$)_{\rm NLTE}\,=\,2.16\,\pm\,0.16$.

We synthesize portions of the CH G-band between 4290 and 4330\,{\AA} 
to determine the C abundance.  
A portion of this region is shown in Figure~\ref{specplot}.
We adopt the CH linelist of B.~Plez (2006, private communication).
Because our C abundance is derived from molecular features
that are very temperature sensitive in a 1D model atmosphere, 
it is likely that the true C abundance is somewhat lower,
as has been found by \citet{asplund05b} when analyzing atomic
and molecular C features in the solar spectrum 
using 3D hydrodynamical model atmospheres.
We cannot estimate the $^{12}$C/$^{13}$C ratio from the CH G-band
at our spectral resolution and $S/N$ levels.

The [O~\textsc{i}] lines at 6300 and 6363\,{\AA}, usually taken as the 
best O abundance indicator in metal-poor stars, are contaminated
by telluric features.
We measure equivalent widths for the O~\textsc{i}
triplet at 7771, 7774, and 7775\AA,
adopt the log($gf$) values from \citet{wiese96}, and 
find $\log\,\varepsilon\,($O$)_{\rm LTE}\,=\,7.56\,\pm\,0.27$.
As summarized in \citet{kiselman01}, the line source function 
for this triplet will show strong departures from LTE,
always in the direction of underestimating the line strength 
and hence overestimating the abundance in LTE relative to non-LTE.
\citet{kiselman01} adopts
a non-LTE correction $\sim\,-0.1$~dex
for solar-type stars with a similar O abundance,
\citet{nissen02} found corrections $\sim\,-0.1$ to $-0.2$~dex
for metal-poor main sequence and subgiant stars, and
\citet{garciaperez06} found corrections $\sim\,-0.1$~dex
for metal-poor subgiant and giant stars.
\citet{garciaperez06} also found that the O abundance derived
from the triplet were $\sim\,+0.1$ to $0.3$~dex higher than when derived from
the [O~\textsc{i}] lines for stars in their sample with a temperature
similar to \ligiant; this difference is also presumably due to 
non-LTE effects.
\citet{cavallo97} found a similar effect in their sample of 
metal-poor subdwarfs and giants, and they reported a difference
of $+0.53$~dex between the triplet and the [O~\textsc{i}] lines.
Adding the two corrections adopted from \citet{garciaperez06}, 
we arrive at a total non-LTE correction for \ligiant\ of 
$\sim\,-0.3$ to $-0.4$~dex.
We therefore derive $\log\,\varepsilon\,($O$)_{\rm NLTE}\,\sim\,7.2$~dex
for \ligiant.

Both lines in the Na~\textsc{i} doublet at 5889 and 5895\,{\AA} are
blended with interstellar absorption lines and telluric emission lines.
The relative radial velocity of \ligiant\ at the time of our
observations does not allow us to separate these components from the
stellar absorption lines.
Instead, we measure equivalent widths from the Na~\textsc{i} lines 
at 5682 and 5688\,{\AA}.
The spectral region around these lines is shown in Figure~\ref{specplot}.
\citet{gratton99} offers non-LTE corrections for these lines 
in metal-poor giants, but \citet{asplund05a} notes that
their corrections
for giants are at odds with other calculations for these lines.
\citet{gehren04} studied the non-LTE abundances of only
dwarfs and early subgiants, so we cannot infer corrections
from their work.
\citet{takeda03} suggest that non-LTE effects for these lines
will be relatively small ($\lesssim\,-0.1$~dex).  

We derive Mg~\textsc{i} abundances from equivalent width measurements
of four lines.
Non-LTE corrections for two of these lines, 4571 and 5183\AA,
have been presented by \citet{gratton99}, who estimate 
corrections
$\sim+0.1$ to $+0.15$~dex in metal-poor giants.
\citet{gehren04} found corrections for dwarfs and subgiants
with similar metallicity to \ligiant\ that are 
$\sim+0.1$~dex.

We measure an equivalent width from the K~\textsc{i} resonance 
line at 7698\AA.
The electron structure of K~\textsc{i} is similar to Na~\textsc{i}, 
and therefore both elements will be sensitive to similar 
non-LTE effects in stellar atmospheres \citep{bruls92}. 
We use the interpolation software kindly provided by 
Y.~Takeda (2007, private communication) to estimate the non-LTE
correction from the analysis performed by \citet{takeda02}.
The correction is rather large, $\sim-0.4$~dex.

We do not adopt any non-LTE corrections for the remaining
six light species that were detected---Ca, Sc, Ti, Cr, Mn, and Ni.

We measure abundances for 10 \ncap\ elements
(Sr, Y, Zr, Ba, La, Ce, Pr, Nd, Sm and Eu) in \ligiant\
by generating synthetic spectra for each line.
Relative wavelengths for the hyperfine structure and isotopic components 
were drawn from
\citet{mcwilliam98} for the Ba~\textsc{ii} 4554\,{\AA} resonance line.
All La~\textsc{ii} lines were synthesized with the hyperfine structure
presented in \citet{lawler01a} and \citet{ivans06}.
Pr~\textsc{ii} lines were synthesized from the hyperfine A constants
presented in \citet{ivarsson01}.
Both Eu~\textsc{ii} lines were synthesized with the
hyperfine structure presented in \citet{lawler01b} and \citet{ivans06}.
Each of these species has notable hyperfine splitting,
and thus it is necessary to synthesize the hyperfine structure
and isotope shifts in order to derive accurate abundances.
Our spectrum of \ligiant\ does not extend far enough into the 
blue to observe the stronger transitions from many of the heavier
\ncap\ species, such as the Pb~\textsc{i} 4057\,{\AA} line.
We note that no Tc is detected at the 4238, 4262, and 4297\,{\AA} 
lines in \ligiant, and we only obtain an upper limit of
$\log\,\varepsilon\,($Tc$) \lesssim +0.5$.

\subsection{Uncertainties}
\label{uncertainties}

Total uncertainties in the derived elemental abundances were estimated
by adding uncertainties in the log($gf$) values, uncertainties in the
EW resulting from continuum placement, 
the statistical scatter associated with
measuring multiple lines of each species, and changes in the derived
abundance in response to our uncertainties in the atmospheric parameters
in quadrature. 
The statistical scatter is $\sim0.1$--0.2~dex for most species
with more than $\sim5$ lines analyzed;
for species with fewer than $\sim5$ lines, the uncertainty 
in the continuum placement for individual lines was found to
contribute $\sim0.1$--0.2~dex to the overall abundance error budget.
For neutral species, the effective temperature uncertainties
translate into abundance uncertainties $\sim0.15$~dex, and
for singly-ionized species the surface gravity uncertainties
translate into abundance uncertainties $\sim0.10$~dex.
For a few species with several strong lines (Mg, Ca, Fe, Ti, and Ni), 
the microturbulent velocity uncertainties translate into abundance 
uncertainties $\sim0.05$~dex.
Uncertainties in the log($gf$) values and overall metallicity
of the model atmosphere result in abundance uncertainties
$\lesssim0.05$~dex.
One extreme case is C, whose abundance was derived from
molecular CH bands; a change in $T_{\rm eff}$ by $\pm$\,150\,K 
changes the C abundance by $\approx\,\pm\,0.30$~dex, which 
dominates over changes in the other atmospheric parameters.

\section{Results}
\label{abundresults}

\subsection{Lithium}
\label{liresults}

Contrary to expectations from the evolved nature of \ligiant,
we have detected the Li 6707\,{\AA} line, deriving an abundance of
$\log\,\varepsilon\,($Li$)_{\rm NLTE} = 2.16\,\pm\,0.16$.
This is consistent with the Li abundance of unevolved metal-poor stars
on the Spite Plateau.
The nearly constant level of Li in stars on the Spite Plateau
is only found for dwarfs and subgiants 
with $T_{\rm eff} \gtrsim 5700$\,K,
which is much warmer than the temperature of \ligiant\
\citep[e.g.,][]{pilachowski93,ryan96,ryan98}. 
The Li-enrichment mechanism of \ligiant\ is likely 
unrelated to warm metal-poor turnoff stars, such as 
CS~22898-027 and LP~706-7.
These stars have metallicities as low as or lower than
\ligiant\ and nearly identical Li abundances, yet they have
$T_{\rm eff}/\log(g) \sim$~6300\,K/4.0 and 6000\,K/3.8, respectively
\citep{thorburn92,norris97,preston01}.
The different evolutionary states of these stars indicate that 
different explanations for the Li enrichment may be necessary.
High levels of Li have been observed in other evolved stars, too, 
but most of these stars have metallicities close to solar.

The surface $^{7}$Li abundance in dwarf stars is 
depleted during main sequence core H-burning.
This is because diffusion by gravitational settling becomes more
efficient as the surface convection zone becomes increasingly
shallow.  Lower-mass ($M\sim0.5$--0.65\,$M_{\sun}$) stars have longer
main sequence lifetimes and also burn more $^{7}$Li per unit time than
higher-mass ($M\sim0.65$--0.75\,$M_{\sun}$) stars due to more 
effective convective mixing in their envelopes.  Therefore the
lower-mass main sequence stars will exhibit more $^{7}$Li depletion
than higher-mass main sequence stars.  As these stars evolve away from
the main sequence their convective zones deepen further, diluting the surface
$^{7}$Li abundance and destroying it in the deeper layers.  
More details of these processes are described in
\citet{deliyannis90} and \citet{proffitt91}.  

\citet{pilachowski93}
showed that Li abundances in metal-poor subgiants continue to decline
by an additional factor of $\sim 10^{1}$--$10^{2}$ relative to the
standard evolutionary model predictions as these stars ascend the RGB
and AGB.  Additional physical mechanisms, such as rotationally-induced
mixing, have been used to explain some of the observed $^{7}$Li
abundance decreases from predictions made from standard 
evolutionary models (see the discussion in, e.g., 
\citealt{pinsonneault92}, \citealt{deliyannis93}, and \citealt{ryan98}).
Even without detailed knowledge of the details of these processes, it
is clear that the derived $^{7}$Li abundance in \ligiant\ cannot be
its zero-age main sequence $^{7}$Li abundance.

There are several reasons to believe that the Li in
\ligiant\ was not transferred from an undetected companion.
Our four radial velocity measurements for this star 
suggest that \ligiant\ is not a member of a binary star system, though 
perhaps the system is a long-period binary or 
the plane of the orbit is nearly face-on with respect to the Earth. 
Even if \ligiant\ is the secondary star of a binary system where
the now-unseen and presumably more massive primary swelled in size
during its TP-AGB phase and transferred mass---including
freshly-synthesized Li---to the secondary, such a
scenario would require extremely efficient mass transfer to enrich
\ligiant\ to $\log\,\varepsilon\,($Li$)\,\sim\,2.1$.

Only two stars with [Fe/H]~$< -1.5$ have been reported to possess
$\log\,\varepsilon\,($Li$) \gtrsim +2.4$: HD~160617 and 
BD$+$23$^{\circ}$3912.
While \citet{charbonnel05} reported $\log\,\varepsilon\,($Li$) = +2.56$
in HD~160617, other studies have found lower abundances in this star,
$\log\,\varepsilon\,($Li$) = +2.23$ \citep{pilachowski93} and
$\log\,\varepsilon\,($Li$) = +2.28$ \citep{asplund06}.
\citet{charbonnel05} also reported $\log\,\varepsilon\,($Li$) = +2.64$
in BD$+$23$^{\circ}$3912, but other studies have also found lower 
abundances in this star,
$\log\,\varepsilon\,($Li$) = +2.38$ \citep{pilachowski93} and
$\log\,\varepsilon\,($Li$) = +2.23$ \citep{thevenin98}.\footnote{
We also note that neither of these subgiants exhibit any significant
overabundances of C or Ba, two key signatures of \spro\ nucleosynthesis.
In HD~160617,
[C/Fe]~$= -0.6$, and [Ba/Fe]~$= 0.0$
\citep{jonsell05,johnson07}.
In BD$+$23$^{\circ}$3912, 
[C/Fe]~$= -0.2$, and [Ba/Fe]~$= +0.1$ 
\citep{fulbright00,gratton00}.}
Thus it appears that Li enrichment significantly above the
Spite Plateau value in metal-poor stars is, at best, a rare 
phenomenon.

Similarly, the lack of Li-enriched stars anywhere above the Spite Plateau
or the Li abundances expected by normal dilution as stars evolve up the 
red giant branch suggests that \ligiant\ was not enriched during
or prior to its departure from the main sequence, in which case
it would be obeying the normal dilution effects predicted
by standard evolutionary models.  
(In other words, the pattern of Li abundances seen in Figure~4b
of \citealt{ryan98} has not been shifted upward by 1--2~dex
due to extrinsic Li enrichment.)

Later, in \S~\ref{gallino}, we consider the possibility that the
\ncap\ material in \ligiant\ was accreted from a companion star
that passed through the TP-AGB phase of evolution.  
In this scenario, we derive a dilution factor of 63 (where
$\log_{10}(63) = 1.8$); i.e., 
one part of accreted material is observable in the stellar atmosphere
of \ligiant\ for every 63 parts of its own, original material 
necessary to produce the observed \ncap\ abundances.
If the Li shares its nucleosynthesis origin with this \ncap\ 
enrichment, it too would be expected to be diluted by the same factor.
Figure~10 of \citet{gratton00} leads us to surmise that the 
un-enriched Li abundance of \ligiant\ would be 
$\log\,\varepsilon\,($Li$) \sim 0$ to 1 at its present evolutionary state.
Therefore, in order to enrich \ligiant\ to a present Li abundance of
$\log\,\varepsilon\,($Li$) = +2.1$, nearly 3--4~dex 
(i.e., the sum of the enrichment and the logarithmic dilution factor)
would need to be acquired by \ligiant\ from its companion.
Furthermore, the transfer would have needed to occur fairly
recently---otherwise the Li acquired by \ligiant\ would be diluted
and destroyed by the normal channels during its evolution up the RGB.
The likelihood of such extreme enrichment seems small.

We conclude that the Li observed in \ligiant\ is not of primordial
origin and was not transferred from an unseen binary companion.
The Li should have originated within this star.

\subsection{CNO, $\alpha$, and Fe-peak Elements}
\label{alphaferesults}

Carbon and oxygen are overabundant in \ligiant, with
[C/Fe]~$\approx +0.7$ and [O/Fe]~$\approx +1.1$.
The CN molecular band at 3850\,{\AA} commonly used to measure the 
N abundance was not covered in our spectra.
Both \citet{mcwilliam95a,mcwilliam95b} and \citet{cayrel04}
found a large amount of scatter in the C abundances of their
very metal-poor star samples, despite the fact that
the \citet{cayrel04} sample is biased against C-rich stars.
It is difficult to ascertain whether the C overabundance
in \ligiant\ is typical for metal-poor stars or is the
result of an additional enrichment process.
If C enrichment has occurred, it is mild 
relative to the population of C-enriched metal-poor stars,
which can have [C/Fe]~$\gtrsim 2.0$ in stars with similar metallicity
to \ligiant\ \citep{cohen06,aoki07}. 
Comparison of our O abundance
with the O abundances of the sample of metal-poor subgiants and giants
from \citet{garciaperez06}, both 
derived from the triplet near 7770\,{\AA},
reveals that the O abundance of \ligiant\ is in very good agreement
with their results.
\citet{garciaperez06} found [O/Fe]~$\sim +0.5$ near [Fe/H]~$=-2.2$
from the [O~\textsc{i}] lines, which is only marginally smaller 
than our O abundance corrected for non-LTE effects, and within the
uncertainties of these measurements the \ligiant\ O abundance is 
not anomalous.
This suggests that no extra enrichment of O-rich material has occurred.
%\citet{cayrel04}, using 1D LTE calculations, found a mean
%O abundance in their sample of metal poor stars,
%[O/Fe]~$=+0.7\,\pm\,0.17$; 
%applying 3D corrections for the dwarfs lowered this abundance 
%by about 0.2~dex.
%Given the large dispersion of their sample, the O abundance
%in \ligiant\ appears typical for metal-poor stars and does 
%not suggest that any extra enrichment has occurred.

\begin{figure}
%\epsscale{0.75}
\epsscale{1.00}
%\plotone{f4.bw.eps}   %% b/w figure, print edition
\plotone{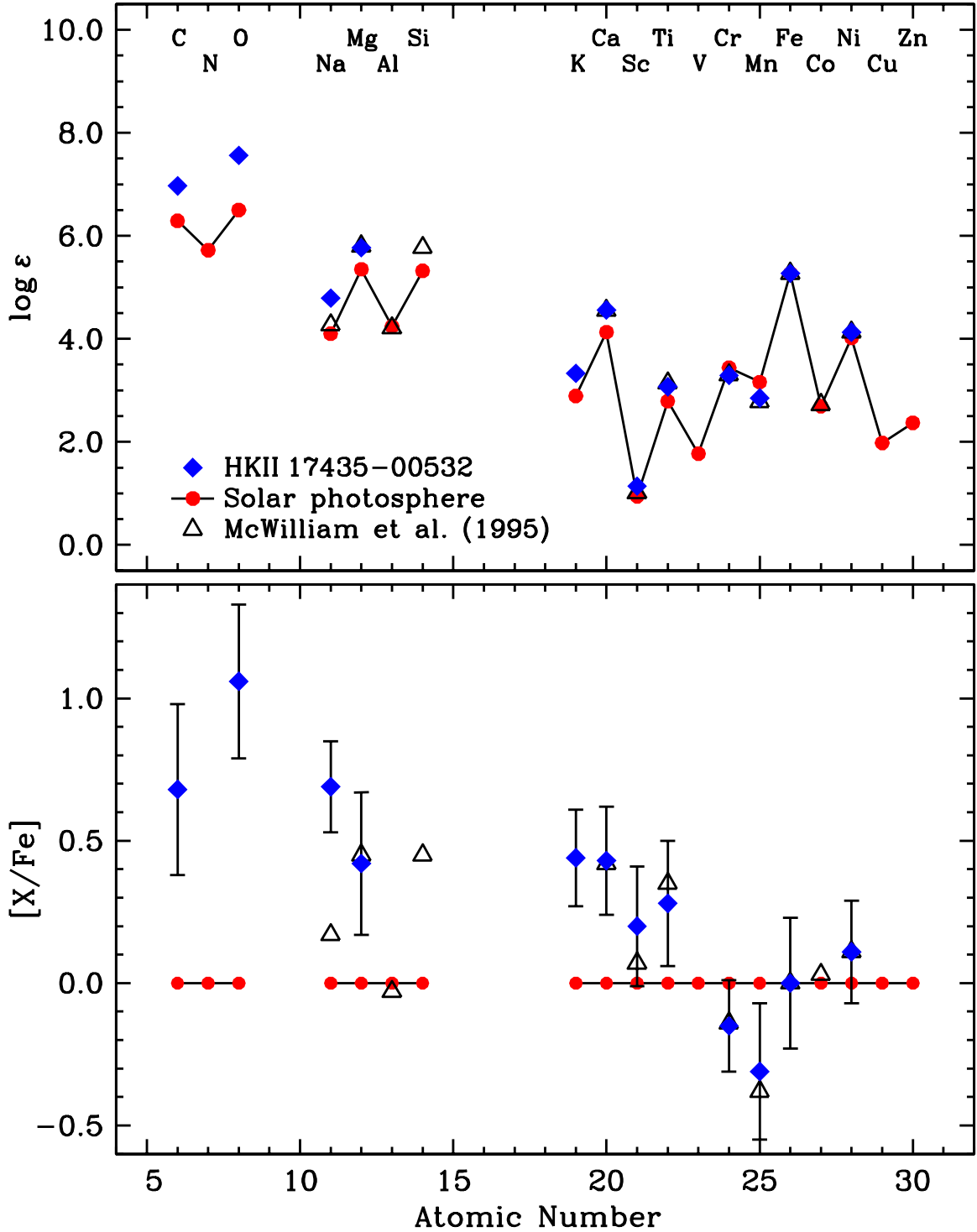}   %% color figure, online edition only
\caption{
\label{alphafe}
Light element ($6 \leq Z \leq 30$) 
LTE abundances for \ligiant, the Sun, and ``typical'' metal-poor stars.
The top panel shows absolute $\log\,\varepsilon$ abundances and 
the bottom panel shows relative [X/Fe] abundances.
We show the abundances for \ligiant\ as filled diamonds 
(blue in the online edition).
We also show the solar photospheric abundances as gray circles
(red in the online edition) and connect them with a solid line.
Open triangles represent ``typical'' metal-poor stars by 
averaging the ten most metal-rich ($-2.8 <$~[Fe/H]~$< -2.0$) stars
in the \citet{mcwilliam95a,mcwilliam95b} sample.
The abundances are normalized to the Fe abundance in \ligiant.
[See electronic edition for a color version of this figure.]
}
\end{figure}

In Figure~\ref{alphafe} we show the elemental abundances for 
$6 \leq Z \leq 30$ in \ligiant, the Sun, and ten ``typical'' metal-poor stars.
We represent the abundances of ``typical'' metal-poor stars by averaging
the ten most metal-rich ($-2.8 <$~[Fe/H]~$< -2.0$) stars
in the \citet{mcwilliam95a,mcwilliam95b} sample.
The well-known odd-even effect is clearly seen in the 
$\log\,\varepsilon$ abundances for these stars in Figure~\ref{alphafe}.
With the exception of Na, the $\alpha$ and Fe-peak elements
appear very typical for a star in this metallicity regime.

Sodium is noticeably overabundant in \ligiant, [Na/Fe]$_{\rm LTE}=+0.69$, 
although inclusion of non-LTE effects for the 5682 and 5688\,{\AA}
Na lines could lower this abundance by $\sim 0.1$~dex.
Abundance analyses of large numbers of metal-poor stars 
\citep[e.g.,][]{mcwilliam95a,mcwilliam95b,cayrel04}
with [Fe/H]~$< -2.0$ have found little change in 
[Na/Fe] for metallicities $-3.0 <$~[Fe/H]~$< -2.0$,
with an intrinsic scatter of a few tenths of a dex.
For the \citet{pilachowski96} field stars that have comparable
temperature and gravity to \ligiant, we note that 
there exists a difference in their LTE Na abundances and 
the LTE Na abundance of \ligiant\ by more than 0.4~dex,
which presumably does represent a physical difference 
in the Na abundances and not, e.g., an effect of 
disregarding non-LTE effects since the same set of Na lines 
were used for the analysis.

\citet{gratton00} found no evidence for a significant change in 
the Na abundance of field stars with $-2.0 <$~[Fe/H]~$< -1.0$
along the RGB, although all of the stars in their sample
exhibited [Na/Fe]~$\leq +0.4$.
\citet{aoki07} also find a number of C enriched metal-poor stars that
have overabundances of Na, including some stars with 
[Na/Fe]~$>+2.0$.
These authors showed that the Na-enhancement in their sample
correlated with C and N enhancements.
We note that the C and Na overabundances in \ligiant\ are
much less extreme than the C and Na overabundances in 
some of the stars in the \citet{aoki07} study.
The available information hints that
the Na overabundance in \ligiant\ may share a common origin with the
C overabundance rather than the evolutionary state of this star.
%Further exploration of this link is beyond the scope of this study.

\subsection{\ncap\ Elements}
\label{ncapresults}

\begin{figure}
\epsscale{1.00}
%\plotone{f5.bw.eps}   %% b/w figure, print edition
\plotone{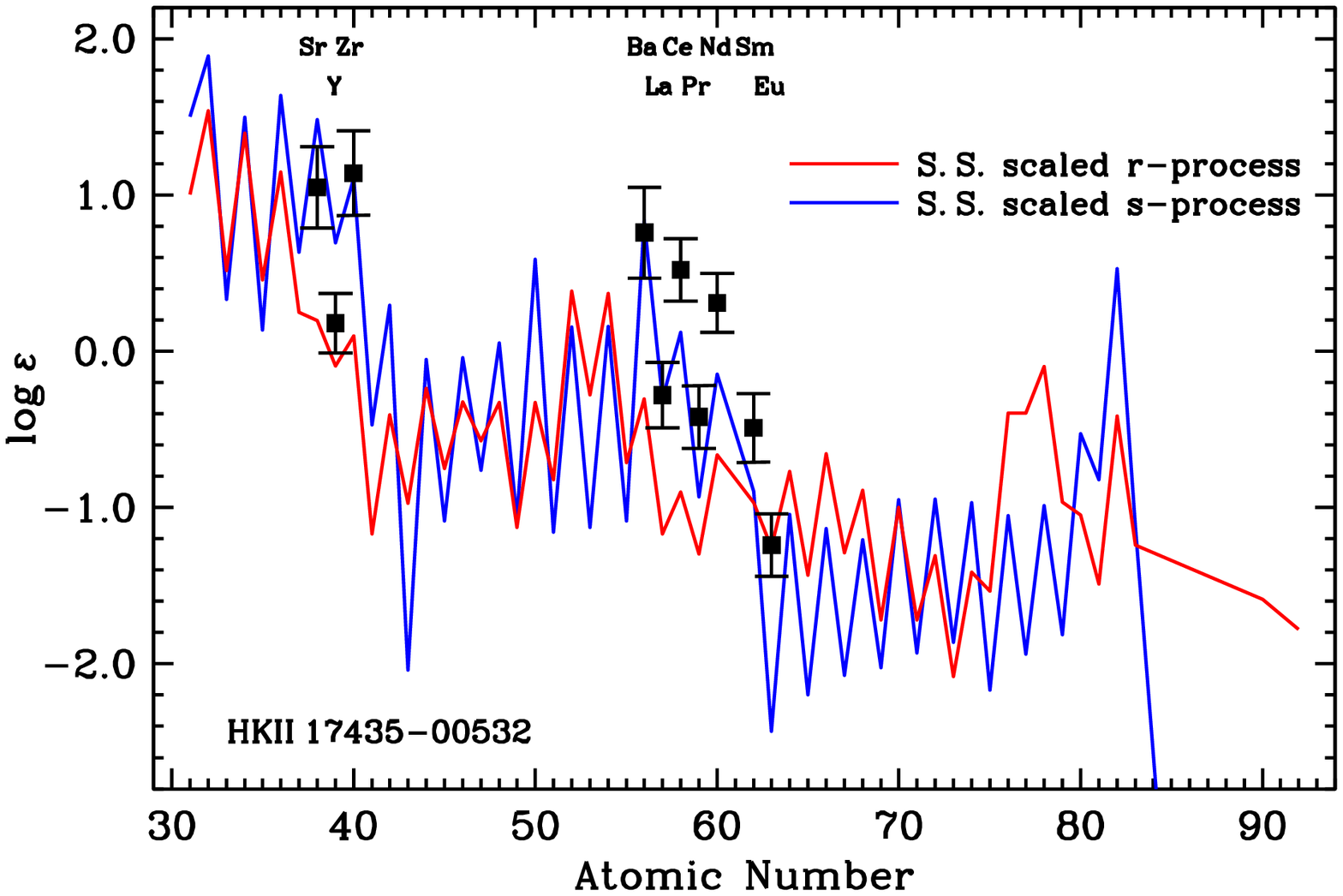}   %% color figure, online edition only
\caption{
\label{ncap}
Derived abundances in \ligiant\
and the scaled S.S.\ $s$- and \rpro\ abundance patterns,
indicated by the gray and black lines, respectively
(blue and red in the online edition).
The S.S.\ \spro\ distribution is normalized to Ba, and
the S.S.\ \rpro\ distribution is normalized to Eu.
It is clear that neither distribution provides a 
satisfactory fit to the \ncap\ elemental abundance pattern
in \ligiant.
[See electronic edition for a color version of this figure.]
}
\end{figure}

\ligiant\ exhibits overabundances of all \ncap\ elements relative
to Fe when compared with the solar system (S.S.) abundance ratios.  
Both Ba and Eu exhibit significant overabundances in \ligiant,
[Ba/Fe]~$=+0.86$ and [Eu/Fe]~$=+0.48$.
These ratios suggest that \ligiant\ is enriched to some degree
in both $s$- and \rpro\ material.
Also, [Ba/Eu]~$=+0.38$,
which suggests that the \spro\ has contributed a greater
portion of the \ncap\ species in this star than the \rpro.
In Figure~\ref{ncap}, we show the derived \ncap\ elemental abundances in 
\ligiant, as well as the scaled S.S.\ $s$- and \rpro\ abundance patterns.
If we examine the entire set of derived \ncap\ abundances in \ligiant,
it is clear that the scaled solar abundances of 
neither process alone provide a satisfactory fit.
Furthermore, within the ranges $38 \leq Z \leq 40$ and $56 \leq Z \leq 58$,
the odd-even effect is more pronounced than would be expected,
with the odd-$Z$ elements exhibiting abundances $\sim$~0.8--1.0~dex
lower than the neighboring even-$Z$ elements.

\begin{figure}
%\epsscale{0.96}
\epsscale{1.00}
\plotone{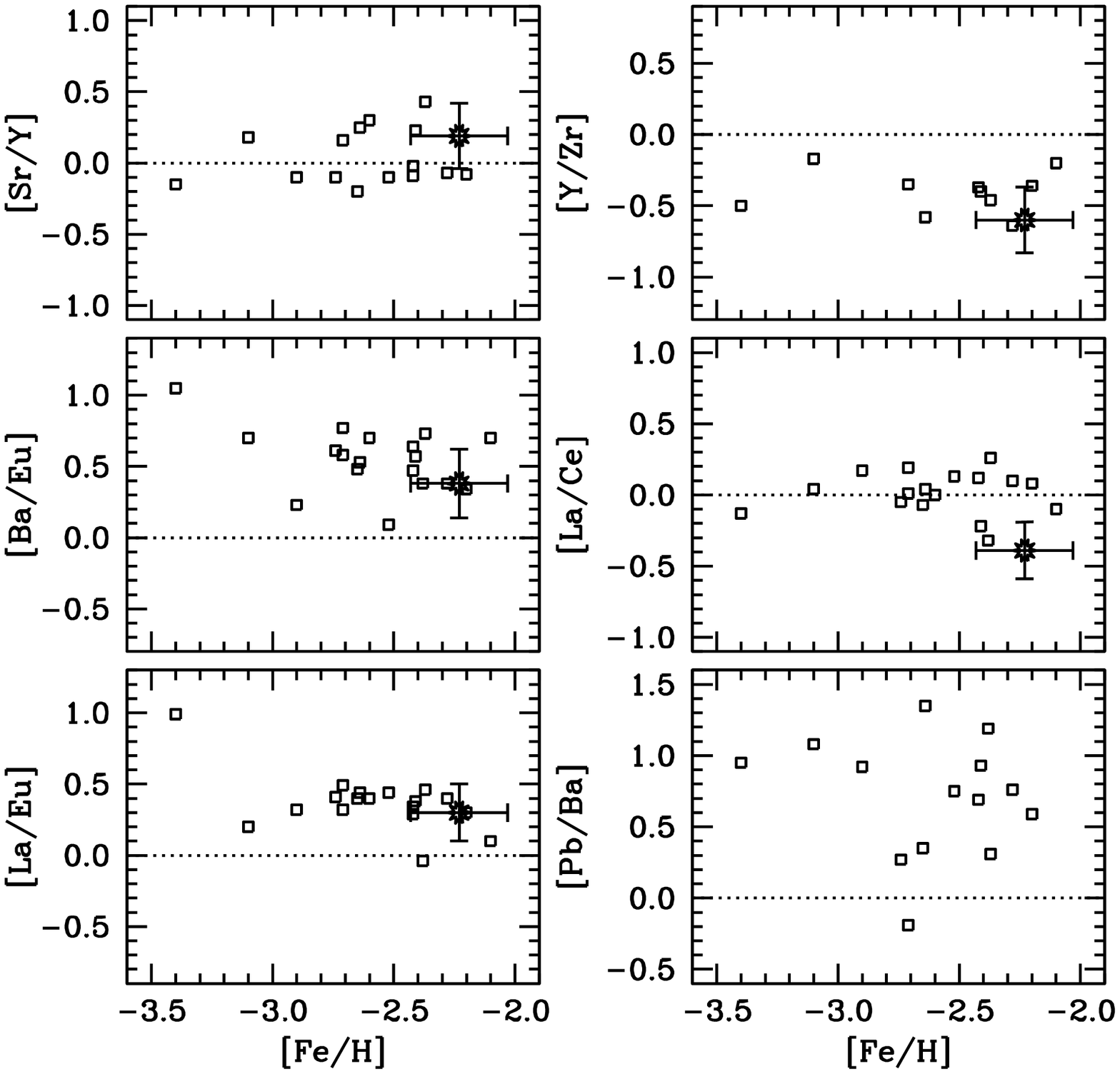}
\caption{
\label{ncap2}
Abundance ratios of \ncap\ species for \ligiant\ and
other stars classified as $(r+s)$-enriched.
\ligiant\ is indicated by the sunburst, and
other stars are indicated by open squares.
The solar photospheric ratios are indicated by dotted lines.
The additional data were taken from other studies 
and the $r+s$ list of \citet{jonsell06}, which was compiled
from the measurements of others, including:
\citet{aoki01},
\citet{aoki02},
\citet{barbuy05},
\citet{barklem05},
\citet{cohen03},
\citet{cohen06},
\citet{hill00},
\citet{ivans05},
\citet{johnson02b},
\citet{johnson04},
\citet{lucatello03},
\citet{masseron06},
\citet{norris97},
\citet{preston01}, and
\citet{sivarani04}.
We also include the $r+s$ star CS~22964-161 \citep{thompson08}.
}
\end{figure}

We display the abundance ratios for several sets of \ncap\ species
in Figure~\ref{ncap2} for \ligiant\ and the 16 stars
classified by \citet{jonsell06} as $(r+s)$-enriched\footnote{
The star CS~31062-012 is also called LP~706-7 and 
has been listed twice in the \citet{jonsell06}
table of $r+s$ stars.
This double identification was also pointed out by \citet{ryan05}.
}$^{\rm ,}$\footnote{
Strictly speaking, the \citet{jonsell06} 
$r+s$ classification refers to stars with 
[Ba/Fe]~$>+1.0$, [Eu/Fe]~$>+1.0$, and [Ba/Eu]~$>0.0$, and 
\ligiant\ does not meet these criteria.
In this paper we refer to $r+s$ stars in a less strict sense 
to indicate that they exhibit some enhancement of
both Ba and Eu, i.e., $+0.3 \lesssim$~[Ba/Fe] and 
$+0.3 \lesssim$~[Eu/Fe].}.
The abundances of the \ncap\ species are all derived 
from transitions of singly-ionized atoms, and the uncertainty 
in the surface gravity of the model atmosphere will have relatively small
effects on the ratios of one \ncap\ species to another.
Several of these ratios were chosen to examine whether the 
exaggerated \ncap\ odd-even-$Z$ effect is common to other stars
enriched in both $s$- and \rpro\ material.
In particular, the [Sr/Y] and [Y/Zr] ratios in \ligiant\ 
appear normal with respect to other stars in this class.
The [La/Ce] ratio is slightly lower in \ligiant\ than the
comparison stars but is still in agreement.
The [Ba/Eu] and [La/Eu] ratios are in agreement.
The [La/Eu] and [La/Ce] ratios exhibit remarkably small scatter 
among the stars of this class with the minor exceptions only for three stars.
One of these stars, CS~29526-110, was noted by \citet{aoki02} 
to have a low La abundance.
\citet{jonsell06} point out that HE~1405-0822 
was noticed to have strong C molecular features in the spectrum
by \citet{barklem05}, and the data presented were preliminary.
CS~30322-023, the most metal-poor star to demonstrate an
\spro\ abundance signature, appears to have a rather low
Eu abundance relative to the second-$s$-process peak elements
\citep{masseron06}.
The exaggerated (relative to the
scaled-S.S.\ $s$- and \rpro\ abundances) 
odd-even-$Z$ effect for $38 \leq Z \leq 40$
and $56 \leq Z \leq 58$ appears to be a characteristic
of the nucleosynthesis in these stars.

Despite the small intrinsic scatter among
these abundance ratios, the scatter in the [Pb/Ba] ratio
is noticeably greater, far beyond any reasonable variation
in the atmospheric parameters or uncertainties in measuring
the abundances of these stars (see also \citealt{masseron06}).
No trend with [Fe/H] is apparent.
It is likely that \ligiant\ has an overabundance of Pb, 
but we do not presently know because
our spectra do not include the Pb~\textsc{i} 4057\,{\AA} 
line to confirm this.
For the often-discussed question of Ba and Pb abundances in the 
($r+s$)-enriched stars we refer to \citet{ivans05}, \citet{bisterzo06}, 
and references therein.

\begin{figure}
%\epsscale{0.61}
\epsscale{1.00}
%\plotone{f7.bw.eps}   %% b/w figure, print edition
\plotone{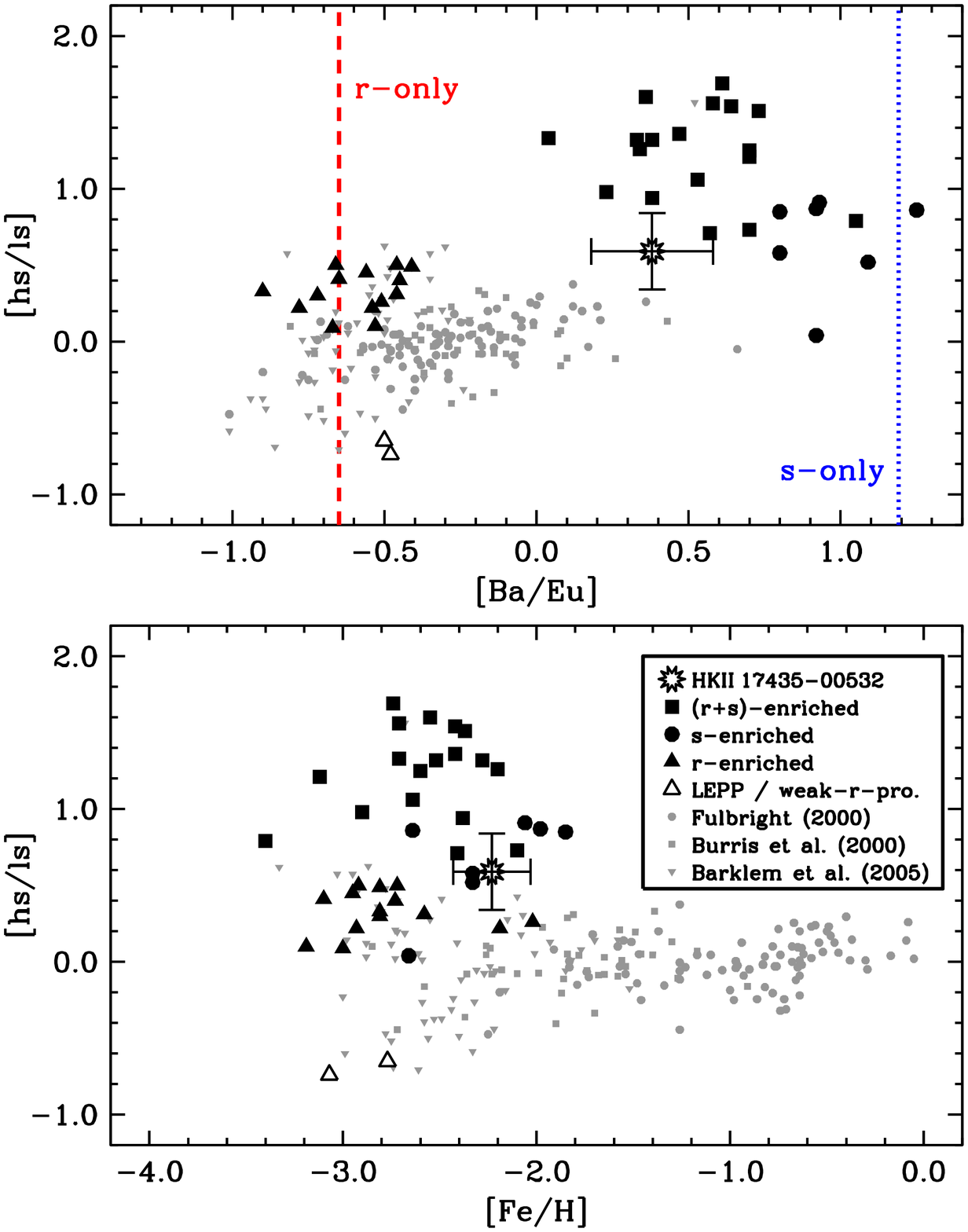}   %% color figure, online edition only
\caption{
\label{lshs}
The [$hs/ls$] abundance ratio is shown as a function of
[Ba/Eu] (top panel) and [Fe/H] (bottom panel).
\ligiant\ is indicated by the sunburst.
Large black squares, circles, and triangles 
indicate stars classified by \citet{jonsell06}
as $(r+s)$-enriched, $s$-enriched, and $r$-enriched, respectively.
Most of the \citet{jonsell06} classifications were based on 
measurements made by 
previous studies of these stars; we supplement their sample
with HD~115444 \citep{westin00},
HD~221170 \citep{ivans06}, 
CS~30322-023 \citep{masseron06}, and
CS~22964-161 \citep{thompson08}.
Open triangles represent HD~122563 and HD~88609 \citep{honda06,honda07}, 
which may be enriched by products of the 
``Lighter Element Primary Process'' (LEPP; \citealt{travaglio04}) 
or the weak-$r$-process \citep{wanajo06,qian07}.
Gray circles, squares, and triangles represent stars from the 
metal-poor abundance surveys of \citet{fulbright00}, 
\citet{burris00}, and \citet{barklem05}.
The stellar model pure-$s$ and pure-$r$ [Ba/Eu] ratios 
\citep{arlandini99} are shown by the dotted line and dashed line
(blue and red in the online edition).
[See electronic edition for a color version of this figure.]
}
\end{figure}

We find a difference in the ratios of the 
mean light-\spro\ abundances 
(Sr, Y, Zr; hereafter designated as $ls$)
and heavy-\spro\ abundances (Ba, La, Ce; hereafter designated as $hs$)
to Fe, 
[$ls$/Fe]~$=0.37\,\pm\,0.13$ and 
[$hs$/Fe]~$=0.96\,\pm\,0.13$.
In Figure~\ref{lshs} we show the [$hs/ls$] ratio for \ligiant\
and other metal-poor stars as functions of [Ba/Eu] and [Fe/H].
Even though the \citet{jonsell06} $r+s$ classification required
[Ba/Fe]~$>1.0$ and [Eu/Fe]~$>1.0$, greater than the [Ba/Fe] and
[Eu/Fe] ratios in \ligiant, the [Ba/Eu] ratios for 
their stars are similar to ours.
This suggests that \ligiant\ may have less overall \ncap\ 
enrichment than the \citet{jonsell06} $r+s$ stars, but the
relative amounts of enrichment may be similar.
Also, \ligiant\ and the $r+s$ stars fall in the range between the
pure-$s$- and pure-$r$ [Ba/Eu] ratios and between the stars classified
as $s$-enriched and $r$-enriched.
The [$hs/ls$] ratio in \ligiant\ may lie between the [$hs/ls$] ratio
in $r+s$ and $r$-enriched stars, but the uncertainties on this
measurement are too great to clearly distinguish between these two classes. 

From the data presented in Figure~\ref{lshs}, it appears that,
for the most part, stars enriched in both $s$- and \rpro\ elements
have high [$hs/ls$] ratios (i.e., [$hs/ls]\gtrsim 0.6$).
Based on the overall chemical homogeneity of the stars
in their $r+s$ class, \citet{jonsell06} argued that the 
$s$- and $r$-enrichment in these stars may not have originated from
independent nucleosynthetic events.
Furthermore, based on the frequency of $r+s$ stars in the HERES
survey \citep{barklem05} relative to the numbers of $r$-II 
and $s$ stars, they concluded that the combination of $s$- and $r$-enrichment
in the same star must point to a common (or at least a dependent
set of) nucleosynthesis event(s).

Lucatello et al.\ (2006) have recently examined the C-rich
and/or very cool stars---some of which include \ncap\
enhancements---that could not be studied by the automated procedures of
\citet{barklem05}. 
Lucatello et al.\ (in prep) are presently considering the
frequency of the various categories of \ncap\ stars in the C-rich/cool
sample, so we defer further discussion of such stars to their study.
We also refer the reader to \citet{bisterzo07} for a thorough 
comparison of chemical yields from stars in the TP-AGB phase 
to the \ncap\ abundances in $r+s$ stars.

Our Tc upper limit in \ligiant, $\log\,\varepsilon\,($Tc$) \lesssim +0.5$, 
is of little use, as it is 2--3~dex greater than the scaled
$s$- and \rpro\ predictions shown in Figure~\ref{ncap}.
%The scaled $s$- and \rpro\ distribution predictions
%in Figure~\ref{ncap} predict 
%$\log\,\varepsilon\,($Tc$)_{s} = -2.0$ and 
%$\log\,\varepsilon\,($Tc$)_{r} = -1.0$.
%The relatively short halflives of Tc isotopes, $\lesssim 10^{6}$\,years,
%immediately eliminate an \rpro\ origin, and the Tc abundance predicted
%by the \spro\ is $\sim$~2.5~dex lower than our upper limit.

Given the metallicity of \ligiant, we can safely surmise that 
this star is very old ($\gtrsim 10$\,Gyr) and has a low mass
($\sim 0.8\,M_{\sun}$).
The $s$ and $r$ nucleosynthesis reactions are not expected
to operate in stars of such low mass,
so they must have been present in the material from which
this star formed or were transferred to it from an 
undetected binary companion.

\section{Interpretation}
\label{interpret}

We arrive at the challenge to find a consistent model to explain
Li-, $s$-, and $r$-process-enhancement, as well as 
C- and Na-enrichment, in a metal-poor star on the RGB, RHB, or early AGB.  
We have argued that the Li must have been produced in this star, 
while the $s$- and \rpro\ elements must have been present in the material
from which the star formed or were transferred to it by a nearby star.
Radial velocity variations have not yet been detected, but we consider 
both binary and single-star explanations
because of the possibility of a long binary orbital period or a
highly-inclined orbit. 

In Table~\ref{rvrs}, we summarize the radial velocity measurements
reported in the literature for \ligiant\ and 18 other $r+s$ stars.
Of these 19 stars , 10 show clear evidence of radial velocity
variations, 5 do not show any 
variations when observed for a substantial period of time (including
\ligiant), and 4 have fewer than two measurements reported in the
literature.  Of the 6 stars with measured orbital periods, 5 have
periods greater than 250 days, and one of these has a period greater
than 12 years.  Given the similarities in the abundance ratios between
these stars and \ligiant, as well as its unproven binary status, 
it is reasonable to expect that this
star may be a long period binary.   Future radial velocity
monitoring will be undertaken to investigate the matter.
A wider binary separation could
also explain the less-extreme overabundances of \ncap\ material in
\ligiant\ relative to other $r+s$ stars, similar to the conclusion
reached by \citet{han95} for the enrichment patterns in Ba and CH
stars in binary systems.  This explanation also gains support from the
work of \citet{lucatello05}, who observed radial velocity variations
in 68\% of their sample of 19 C and $s$-process-enriched stars, 
yet were able to conclude from their Monte Carlo analysis of the sample 
that all C- and $s$-enriched metal-poor stars should be members of 
binary systems.

\subsection{Lithium Self-Enrichment through Extra Mixing}
\label{intrinsicli}

\subsubsection{On the RGB}
\label{lirgb}

The first dredge-up episode occurs on the lower RGB
when the outer convective envelope deepens and encounters
material that has been processed through H-burning.
Because it mixes fresh H from the surface downward into the star,
the first dredge-up leaves a strong molecular weight
discontinuity at the point of maximum penetration into the star.
This material is outside of the H-burning shell, and when the
H-burning shell burns outward through this molecular weight discontinuity
it encounters a fresh supply of H.
As a result, the outward progress of the H-burning shell is halted,
the radius of the star ceases to expand, and the 
star's ascent up the RGB is paused.
The RGB luminosity bump occurs at the point on the RGB where
stars spend a larger fraction of their time digesting this H fuel.
The location of the RGB luminosity bump (which depends on metallicity), 
inferred from the calculations of \citet{denissenkov03}, is 
indicated in Figure~\ref{hrd}. 

Before the H-burning shell erases the molecular weight discontinuity,
no extra mixing can occur; 
after it has been erased, the star continues to evolve up the RGB.
To explain the Li enrichment observed in some evolved stars,
several authors \citep{charbonnel95,sackmann99,charbonnel00}
proposed that an extra, yet unspecified, mixing process
may occur in low- and intermediate-mass stars 
(i.e., $1\,\lesssim\,M\,\lesssim\,5\,M_{\sun}$)
after the molecular weight discontinuity has been erased 
at the RGB luminosity bump.
The 3D simulations of the He-flash presented in \citet{dearborn06}
also suggest that this extra mixing may be present outside the 
H-burning shell. 
This extra mixing could drive $^{7}$Li production by the 
\citet{cameron71} mechanism.
This hypothesis explains the low- and intermediate-mass
Li-enriched stars found at the RGB luminosity bump.

Once the H-burning shell begins to exhaust its fresh supply of H
it continues its outward burn, and 
the star once again expands and continues to ascend the RGB.
The convective envelope extends deeper into 
the star, carrying the $^{7}$Li to these hotter regions 
where it is destroyed.
The Li-rich phase is fleeting.
From the calculations presented in \citet{denissenkov04} 
for Li-enrichment in low-mass, metal-poor stars along the RGB, 
we estimate that the Li-rich phase will not last more than 3--4\,Myr.

What is the physical cause of the extra mixing process, and
can it operate in low-mass, low-metallicity stars?
Thermohaline mixing \citep{ulrich72,kippenhahn80,eggleton06,eggleton08}
has long been considered plausible.
This type of mixing is a result of a mean molecular weight inversion
that arises in stars on the AGB when the 
$^{3}$He($^{3}$He,\,2p)$^{4}$He reaction lowers the mean
molecular weight (from 3 to 2 for the species that
participate in this reaction) near the upper boundary 
of the H-burning shell. 
\citet{charbonnel07} advocate a mechanism that relies on the 
double-diffusion of both the mean molecular weight inversion 
and temperature instabilities to induce mixing. 
They successfully link changes in the surface $^{7}$Li abundance 
(as well as a decrease in C and $^{12}$C/$^{13}$C
and an increase in N) to this extra mixing mechanism
in low-mass ($M\sim0.9\,M_{\sun}$), 
low-metallicity ([Fe/H]~$=\,-$1.8, $-$1.3, $-$0.5) stars 
at and above the RGB luminosity bump.

% from Tim: mention this stuff somewhere here:
%page 20, para 3 -- I would also mention the recent work of Aoki et al. (2007),
%whose measurements of MSTO stars from SDSSS/SEGUE argue AGAINST the operation
%of thermohaline mixing, and also the work (on astro-ph) of Denissenkov &
%Pinsonneault (arXiv:0709.4240) which provides the theoretical discussion for
%lack of the operation of thermohaline mixing (and other pertinent info).

Another possibility, known as the Li flash \citep{palacios01},
which was found as a possible source of Li enrichment in
solar-metallicity, 1.5\,$M_{\sun}$ stars at the RGB luminosity bump.
The creation of a thin Li burning shell induces a convective 
instability, which carries Li to the surface of the star.
The radius and luminosity of the star also increase, lifting the
star off the RGB luminosity bump on the HR diagram.
If such a process could occur in low-mass, low-metallicity stars,
this could also explain the Li enrichment observed in \ligiant.

If \ligiant\ is ascending the RGB, the luminosity of this star
is coincident with the luminosity of the bump for [Fe/H]~$=-2.2$.
It is possible that the present Li abundance has already been
reduced from its maximum abundance.
Since we cannot measure the $^{12}$C/$^{13}$C ratio from our
spectrum, we have no diagnostic tool to infer whether the surface Li
abundance is still increasing, holding steady, or already decreasing. 
Such information would link the evolutionary status of \ligiant\ 
with the apparent self-enrichment with Li:
once the extra mixing extends deep enough to convert 
$^{12}$C to $^{13}$C, the hotter temperatures there will
begin to destroy Li \citep{gratton00,spite06}.
If \ligiant\ is evolving through the RGB luminosity bump,
it is the most metal-poor star for which an extra mixing 
mechanism has been shown to produce Li enrichment in the stellar envelope
at this phase of evolution.

\subsubsection{On the RHB or Early AGB}
\label{liabg}

Enrichment of Li is not predicted to occur on the RHB or early AGB.
\citet{charbonnel00} note that an analog of the RGB extra mixing
would have to be extremely efficient to connect the 
$^{3}$He-rich envelope and the H-burning shell in lower-mass, 
lower-metallicity stars ascending the early AGB.
The second dredge-up, which occurs after core-He-burning has ceased,
does not affect the surface composition of stars with 
$M \lesssim 4\,M_{\sun}$ \citep{karakas02}.
If our adopted surface gravity is too high and our adopted temperature
is too cool, the star would still lie on the RHB or early AGB
and---although we may have overestimated the mass of this star by
$\sim 0.1$--$0.2\,M_{\sun}$---the Li question would remain.

We note that for stars in the later TP-AGB stage,
H-burning will alter the surface composition---including
Li abundance---when the outer
convective zone overlaps with the H-burning shell
(hot bottom burning, or HBB; e.g.,
\citealt{sugimoto71,forestini97} and references therein) 
or when an extra mixing mechanism in the radiative layer above
the H-burning shell connects it to the convective zone
(cool bottom processing; e.g.,
\citealt{boothroyd95,wasserburg95,nollett03} and references therein).
These processes occur in higher-luminosity stars 
(i.e., $\log(L/L_{\sun}) \gtrsim 3$), not lower-luminosity stars
near the base of the AGB \citep[e.g.,][]{kraft99,dominguez04}, 
and they are not suspects
for the extra mixing in \ligiant.
(Recall that $\log(L/L_{\sun}) \approx 2.0$ for \ligiant.)

If \ligiant\ has recently arrived on the RHB from the tip of the RGB,
is it possible that the supposed large Li overabundance that was
produced by the \citet{cameron71} mechanism at the RGB luminosity bump
has not yet been fully-depleted?  
If we assume that the Li-rich phase will not last more than
3--4\,Myr, this timescale is far shorter (by nearly two orders of
magnitude) than the timescale necessary for the star to move from the
RGB luminosity bump to the RHB \citep[e.g.,][p.\ 252]{campbell07}.
We therefore dismiss this hypothesis. 

Also, for metal-poor stars in globular clusters on the RGB or AGB, 
we note that Li production does not need to be associated with C, Na, 
\rpro\, or \spro\ enrichment \citep[e.g.,][]{kraft99,kraft00}.

If \ligiant\ is on the RHB or early AGB, we are left to
postulate that a previously-unidentified 
efficient extra mixing episode may be operating during this
stage of evolution in low-mass, low-metallicity stars.

\subsection{Enhanced Sodium Abundance}
\label{na}

Several $r+s$ stars, including \ligiant, exhibit enhanced Na abundances,
which we summarize in Table~\ref{naabund}.
%\citet{mowlavi99} showed that 
$^{23}$Na can be produced in thermally-pulsing stars on the AGB through
a series of reactions that convert CNO material produced
during He-burning into $^{23}$Na.
Briefly, as summarized by \citet{sneden08},
the H-burning shell of an AGB star first converts CNO nuclei
into $^{14}$N.
Then in the early phases of a thermal instability $^{22}$Ne is generated
in the $\alpha$-capture reaction chain
$^{14}$N($\alpha$,$\gamma$)$^{18}$F($\beta^+\nu$)$^{18}$O($\alpha$,$\gamma$)$^{22}$Ne.
Finally, enhanced $^{23}$Na is produced by \ncap\ on the abundant 
$^{22}$Ne and subsequent $\beta^{-}$ decay.
The available CNO nuclei include those present at the star's birth and
fresh
``primary'' $^{12}$C mixed into the envelope by previous 
third dredge-up (TDU) episodes.
This primary source becomes predominant in very
metal-poor AGB stars (e.g., \citealt{gallino06}).
After the last TDU it is predicted that [Na/Fe]~$\sim$~[Ne/Fe]
(with Ne almost pure $^{22}$Ne) in the AGB envelope.

\subsection{$r+s$ Enrichment}
\label{rsenrichment}

Unlike the case of Li-enrichment, 
these scenarios are virtually independent of the evolutionary state, 
only assuming that this star is not in the TP-AGB stage.

\subsubsection{Pre-Enrichment}
\label{rspre}

The simplest explanation assumes that 
\ligiant\ is---and always has been---a single star.
This scenario assumes that the material from which 
\ligiant\ formed was enriched in both $s$- and \rpro\ material
and that no additional \ncap\ enrichment occurred.
Evidence for \spro\ enrichment has been observed in 
a few stars with very low metallicities
(e.g., CS~22183-015: [Fe/H]~$=-3.1$, \citealt{johnson02b};
CS~29497-030: [Fe/H]~$=-2.8$, \citealt{sivarani04};
CS~30322-023: [Fe/H]~$=-3.4$, \citealt{masseron06}), 
and \rpro\ material has been found in stars of all metallicities
(at least for those with [Fe/H]~$\gtrsim -4.0$).
In a generation of stars that preceded \ligiant, it is possible
that at least one binary pair formed with one star that would ultimately 
explode as a supernova, producing \rpro\ material, and another
star that would ultimately produce \spro\ material during its
TP-AGB phase.
This $s$- and $r$-enriched material would then be recycled into the
ISM from which \ligiant\ later formed.
Such a scenario cannot be ruled out and would also 
account for the possible link between the $s$ and $r$ enrichment
mechanisms.

\subsubsection{Pollution from a 1.5\,$M_{\sun}$ Companion}
\label{gallino}

\begin{figure}
\epsscale{1.00}
%\plotone{f8.bw.eps}   %% b/w figure, print edition
\plotone{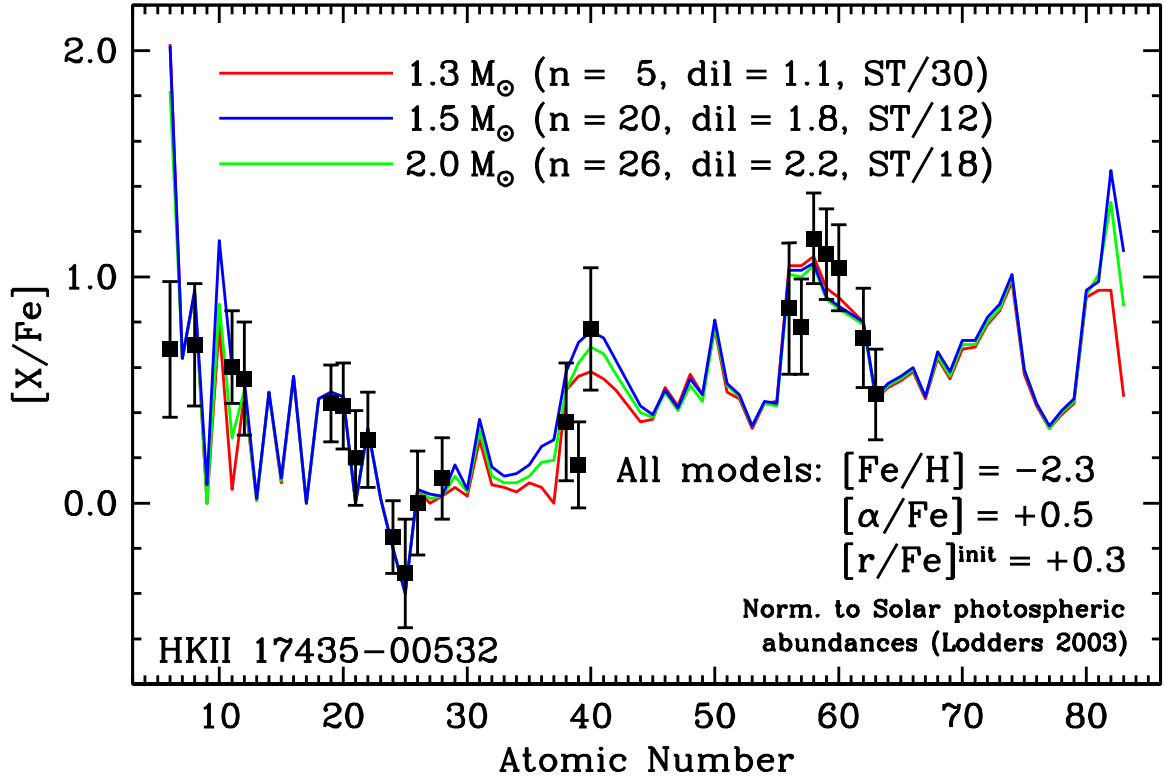}   %% color figure, online edition only
\caption{
\label{gallinoplot}
Predicted [X/Fe] ratios in \ligiant\ assuming pollution from a companion
star that passed through the TP-AGB phase.
The abundance ratios have been normalized to the 
solar photospheric values given in \citet{lodders03}.
LTE abundances are displayed for all elements except O, Na, and Mg,
for which the non-LTE corrected abundances are displayed.
The asymmetric uncertainties on Na reflect a conservative range
on the non-LTE corrections.
All sets of abundance predictions assume [Fe/H]~$=-2.3$, 
[$\alpha$/Fe]~$=+0.5$, and [$r$/Fe]$^{\rm init}=+0.3$.  
The initial mass of the TP-AGB star is the primary variable between
the different sets of abundance predictions, although changing the 
mass necessitates altering the number of thermal pulses (``$n$''), 
the logarithmic dilution factor (``dil''), and the $^{13}$C pocket 
efficiency (``ST/'').
The black curve (blue in the online edition) reflects our best fit model, with
$M_{\rm AGB}=1.5\,M_{\sun}$, while the dark gray curve 
(red in the online edition) and the light gray curve 
(green in the online edition) have
$M_{\rm AGB}=1.3\,M_{\sun}$ and $M_{\rm AGB}=2.0\,M_{\sun}$, respectively.
[See electronic edition for a color version of this figure.]
}
\end{figure}

Let us assume the binary scenario for \ligiant, in which we consider
the observed star to be the longer-lived secondary.
We calculate the \spro\ yields from the primary,
assuming that both members of the binary had the same
initial metallicity.
These calculations were based on the FRANEC stellar evolution models
\citep{straniero97,straniero03,zinner06}
for low-metallicity stars on the AGB.
The best-fit is obtained with the model that adopts an initial mass
of $1.5\,M_{\sun}$ for the companion star that passed through the 
TP-AGB phase.  
Abundance predictions for three sets of model parameters are displayed
in Figure~\ref{gallinoplot}.
The initial mass of the companion star is the
independent variable, although changing the mass affects the number of
$s$-process-producing thermal pulses (``$n$'') that occur,
the logarithmic dilution factor (``dil''; defined to be the logarithm 
of the ratio of original mass in the observed stellar envelope to the 
amount of mass that is acquired), and the $^{13}$C pocket efficiency 
(``ST/''; i.e., some fraction of the standard $^{13}$C pocket efficiency 
of \citealt{gallino98}).  As basic model input, we adopt [Fe/H]~$=-2.3$, 
[$\alpha$/Fe]~$=+0.5$, and [$r$/Fe]$^{\rm init}=+0.3$.

If we adopt the various Na non-LTE corrections for the 5682 and 
5688\,{\AA} lines at face value, a conservative range for the Na 
abundance is found to be $+0.45 <$~[Na/Fe]~$< +0.85$.
This constrains the mass of the undetected companion to 
$1.3 \lesssim M \lesssim 2.0\,M_{\sun}$.
Of the elemental abundances we have measured in \ligiant, the Na
abundance is the best discriminator of the mass of the companion star.
The $\alpha$ and Fe-peak abundance ratios are insensitive to the
model parameters.  

We derive a dilution factor of 1.8~dex (i.e., a factor of 
$\approx 63$) for this system, assuming the companion had a 
main sequence mass of $1.5\,M_{\sun}$. 
An initial enrichment of \rpro\ material is assumed in these models,
but Eu is the only species that we measure in \ligiant\ that is
strongly sensitive to this enrichment.
If we assume an initial [$r$/Fe]~$=+0.3$ for this system, then 
only $\sim 0.2$~dex of Eu needs to be acquired from the 
\spro\ material of the AGB companion.  
In contrast to the Eu, which mostly reflects the initial composition
of the ISM from which the system formed, the Ba 
(and other $hs$ species with large overabundances) was dominantly 
produced by the companion star during its TP-AGB phase and transferred
to \ligiant\ by \spro\ enriched winds.

Predictions for the abundances of \ncap\ species vary somewhat depending
on the parameter choices, but they generally agree with the observations.
The most notable exception is Y, which is predicted to be overabundant
by $\sim +0.5$ to $+0.7$~dex but is measured to have 
[Y/Fe]~$=+0.17\,\pm\,0.19$.
Similar results were found by \citet{aoki06} for CS~31062-050 and LP~625-44.
If we were to adopt initial enrichment levels for Sr, Y, and Zr from the
distribution of unevolved stars compiled by \citet{travaglio04}, 
[Sr/Fe]~$=+0.2$, [Y/Fe]~$=-0.2$, and [Zr/Fe]~$=+0.4$, the changes in the
final abundance predictions for [Y/Fe] are only decreased by 
$\approx 0.1$~dex, while the predicted [Sr/Fe] and [Zr/Fe] ratios
remain approximately unchanged.

\subsubsection{Enrichment from an 8--10\,$M_{\sun}$ Companion}
\label{highmass}

%Variations of the scenario described in \S~\ref{gallino}
%attribute both the $s$- and \rpro\ enrichment to the {\textit same}
%primary star in the binary system.
Another enrichment scenario invokes a single high-mass companion
star as the donor of both the $s$- and \rpro\ material to \ligiant.
Type~II supernovae with $8.0 \lesssim M \lesssim 10.0\,M_{\sun}$
have been identified as a possible source of \rpro\ elements
(specifically, Eu) in low-metallicity stars
\citep{mathews92,ishimaru99,wanajo03,ishimaru04,ning07}.
Similarly, at progressively lower metallicities, stars with smaller 
initial masses will end the AGB phase of their life with
masses above the Chandrasekhar limit and may explode as
Type~1.5 (or AGB) supernova and produce
\rpro\ material \citep{iben83,zijlstra04}.
It has been shown \citep[e.g.,][]{gallino98,busso99} that the
main component of the \spro\ occurs during the interpulse phases
of intermediate-mass stars on the AGB ($1.5 < M < 3.0\,M_{\sun}$).
Due to the accelerated evolutionary rates of stars with 
$M > 3.0\,M_{\sun}$ and the relatively small amount of mass located
in the intershell region, though, very small amounts of \spro\ material
are expected to be produced by TP-AGB stars in the 
$8.0 \lesssim M \lesssim 10.0\,M_{\sun}$ range.
Consequently we dismiss this unlikely enrichment scenario for 
\ligiant.

\subsubsection{Enrichment from Two Companion Stars}
\label{triplestar}

A long-period triple star system could be invoked to explain
\rpro\ material from a high-mass 
($M \gtrsim 8\,M_{\sun}$) supernova and \spro\ material
from an intermediate-mass ($1.5 \lesssim M \lesssim 3.0\,M_{\sun}$)
star on the AGB,
both of which deposited some of this $n$-capture-enriched material on the
tertiary star.
In the case that such a system is dynamically stable 
for the length of the lifetime of a high-mass star, the object could evolve, 
explode as a supernova, and enrich the two remaining stars with 
\rpro\ material.
If the two remaining stars then became separated from the 
compact supernova remnant but remained gravitationally bound,
the intermediate-mass star could evolve, pass through the 
TP-AGB phase, and transfer $s$-process-enriched material to the
low-mass tertiary.  
%Such systems should not form often enough to account for the
%observed frequency of $r+s$ objects. 
The very large orbital separation that would be necessary to maintain
the triple system for a long period of time would necessarily
imply a large dilution of the transferred material.
Though perhaps a less-likely scenario than the single 1.5\,$M_{\sun}$
AGB companion scenario, the triple star system nevertheless could
account for the \ncap\ enrichment pattern observed in \ligiant.

\section{Conclusions}
\label{conclusions}

We have performed the first detailed abundance analysis on \ligiant, a
star that was identified as a metal-poor candidate during the original
HK survey and re-identified during the HK-II survey.
High-resolution spectra necessary for this abundance analysis
were obtained with the HRS on the HET at
McDonald Observatory. No radial velocity variations are detected over
a time span of $\sim180$~days, which prevents us from distinguishing 
whether \ligiant\ is a single star, a member of a long-period binary system, 
or in a binary system with a highly-inclined orbit.  Future radial velocity
monitoring may shed light on the situation.  We derive $T_{\rm
eff}=5200\,\pm\,150$\,K and $\log(g)=2.15\,\pm\,0.4$ for \ligiant,
which places it on the RGB or the RHB/early AGB.  Our abundance
analysis reveals that this star is indeed metal-poor, with
[Fe/H]~$=-2.2$.

A surprising result of our study is the high Li abundance, 
$\log\,\varepsilon\,($Li$)=+2.1$, which is not predicted by
standard stellar models for a low-mass, low-metallicity, evolved star
such as \ligiant.  If this star is evolving up the RGB, it is located
near the RGB luminosity bump, where extra mixing has been found to
cause a short-lived phase of Li enrichment in more massive stars.  In this
case, \ligiant\ would represent the most metal-poor object known for
which this extra mixing phenomenon occurs.  If the star is on the RHB
or early AGB, any Li that was present at earlier times should have
been destroyed.  In this case, we propose that an analog of the extra
mixing that occurs at the RGB luminosity bump may be occurring on the
RHB/early AGB. Future theoretical analysis of this point would be
necessary to confirm or refute this hypothesis.
Measurement of the $^{12}$C/$^{13}$C ratio would also help to better
constrain the evolutionary state of this star.

Our abundance analysis also finds that C and Na are slightly enhanced.
The abundances of O, the $\alpha$-elements, and the Fe-peak elements
all appear typical for stars with metallicity similar to \ligiant.
This star is enriched with both $s$- and \rpro\ elements.
Several scenarios can be invoked to explain the \ncap\ abundance patterns,
as well as the C and Na enhancements.
These include pre-enrichment from supernovae, mass-transfer from a
companion in the TP-AGB phase, or enrichment from a nearby supernova.
%If the $s$ and $r$ enrichment episodes that create $r+s$ stars 
%are not stochastically-independent, then we favor scenarios in which
%a single high-mass star enriched its local environment with 
%both \spro\ and \rpro\ material (from its evolution through the TP-AGB
%phase and during its supernova explosion, respectively).
%This high-mass star could either
%have pre-enriched the ISM from which \ligiant\ formed or could
%remain today as an undetected companion to \ligiant.
Our favored scenario involves pollution from a binary companion star
with an initial mass of $\approx 1.5\,M_{\sun}$; this would only require 
that a small initial enrichment of \rpro\ material was present in the ISM
from which this star formed.

Inefficient mixing of the ISM, differing efficiencies of 
stellar winds from stars on the AGB, or a wide range of orbital separations
can likely account for the range of 
C, Na, $s$-, and \rpro\ enrichments observed in $r+s$ stars.
Future studies of the frequency, binarity, and abundance patterns of stars
enriched in both $s$- and \rpro\ material will be necessary to 
understand the origins of this class of stars.

\acknowledgments

We gratefully thank 
Amanda Karakas and 
Oscar Straniero for enlightening discussions
and the referee for providing a helpful set of constructive suggestions 
on the manuscript.
We also acknowledge
Yoichi Takeda for sharing his K non-LTE correction interpolation routine, 
Eusebio Terrazas and Frank Deglman 
for their support of these HET observations,
and Joy Chavez for taking one of the epochs of the 2.7\,m data.  
We are indebted to Taylor Chonis and Martin Gaskell for obtaining new
photometry of \ligiant\ from the Miller Observatory.
The Hobby-Eberly Telescope (HET) is a joint project of the University 
of Texas at Austin, the Pennsylvania State University, Stanford University, 
Ludwig-Maximilians-Universit\"{a}t M\"{u}nchen, and 
Georg-August-Universit\"{a}t G\"{o}ttingen.
The HET is named in honor of its principal benefactors, 
William P.\ Hobby and Robert E.\ Eberly.
This research has made use of the 
NASA Astrophysics Data System (ADS),
NIST Atomic Spectra Database,
The Amateur Sky Survey (TASS), 
Two Micron All-Sky Survey (2MASS), 
and SIMBAD databases.
The reliability and accessibility of
these online databases is greatly appreciated. 
A.~F.\ acknowledges support through the W.~J.~McDonald Fellowship of
the McDonald Observatory.
J.~R. gratefully acknowledges partial research support for this work by
NASA through the AAS Small Research Grant Program and the GALEX GI grant
05-GALEX05-27.
R.~G. acknowledges the Italian MIUR-PRIN06 Project ``Late phases of 
Stellar Evolution: Nucleosynthesis in Supernovae, AGB Stars, Planetary 
Nebulae'' for support.
Funding for this project has also been generously provided by 
the U.~S.\ National Science Foundation
(grant AST~06-07708 to C.~S;
grants AST~04-06784, AST~07-07776, and PHY~02-15783 to T.~C.~B.\ and
the Physics Frontier Center/Joint Institute for Nuclear Astrophysics (JINA);
and AST~07-07447 to J.~J.~C.).

%% Comment these lines when compiling on IUR's machine!
%
%{\it Facilities:} 
%\facility{HET (HRS)
%\facility{Smith (2dCoud\'{e})}

\clearpage

\begin{deluxetable}{ccc}
\tablecaption{Summary of Radial Velocity Measurements 
\label{rvdata}}
\tablewidth{0pt}
\tablehead{
\colhead{HJD } &
\colhead{RV (km\,s$^{-1}$) } &
\colhead{Facility } }
\startdata
2454140.79715 &  38.61  (0.64) &  HET$+$HRS         \\
2454159.73386 &  38.77  (0.57) &  HET$+$HRS         \\
2454297.61639 &  39.36  (0.59) &  McD2.7m$+$cs21    \\
2454318.60116 &  38.79  (1.07) &  McD2.7m$+$cs23    \\
\enddata
\end{deluxetable}

\clearpage

\begin{deluxetable}{ccc}
\tablecaption{Basic Stellar Data and Model Atmosphere Parameters
\label{basicdata}}
\tablewidth{0pt}
\tablecolumns{3}
\tablehead{
\colhead{Quantity} &
\colhead{Value} &
\colhead{Source} }
\startdata
R.~A.\ (J2000.0)       & $11^{h}49^{m}03.3^{s}$ & 1 \\ 
Dec.\ (J2000.0)        & $+16^{\circ}58'41.7''$ & 1 \\
$V$                    & 13.145$\,\pm\,$0.166   & 2 \\
$I$                    & 12.237$\,\pm\,$0.067   & 2 \\
$J$                    & 11.534$\,\pm\,$0.021   & 3 \\
$H$                    & 11.121$\,\pm\,$0.023   & 3 \\
$K$                    & 11.049$\,\pm\,$0.020   & 3 \\
$E(B-V)$               & 0.042                  & 4 \\
$\log(L)$ ($L_{\sun}$) & 2.01$\,\pm\,$0.50      & 1 \\
$T_{\rm eff}$ (K)      & 5200$\,\pm\,$150       & 1 \\
$\log(g)$              & 2.15$\,\pm\,$0.4       & 1 \\
$v_{t}$ (km\,s$^{-1}$) & 2.0$\,\pm\,$0.3        & 1 \\
${\rm [M/H]}$          & $-$1.85$\,\pm\,0.23$   & 1 \\
${\rm [Fe/H]}$         & $-$2.23$\,\pm\,0.23$   & 1 \\
\enddata
\tablerefs{
(1)~this study;
(2)~TASS;
(3)~2MASS;
(4)~\citet{schlegel98}}
\end{deluxetable}

\clearpage

\begin{deluxetable}{cccccc}
\tablecaption{Equivalent Width Measurements
\label{ewmeasure}}
\tablewidth{0pt}
\tablehead{
\colhead{species} &
\colhead{$Z$} &
\colhead{$\lambda$ (\AA)} &
\colhead{E.P.\ (eV)} &
\colhead{$\log(gf)$} &
\colhead{E.W.\ (m\AA)} 
}
\startdata
Na~\textsc{i}    & 11  & 5682.63   &   2.100  &   $-$0.699  &     20.0 \\
Na~\textsc{i}    & 11  & 5688.21   &   2.100  &   $-$0.456  &     33.0 \\
Mg~\textsc{i}    & 12  & 4571.10   &   0.000  &   $-$5.393  &     73.7 \\
Mg~\textsc{i}    & 12  & 4702.99   &   4.330  &   $-$0.380  &    113.9 \\
Mg~\textsc{i}    & 12  & 5183.62   &   2.720  &   $-$0.158  &    233.2 \\
Mg~\textsc{i}    & 12  & 5528.42   &   4.330  &   $-$0.500  &     87.0 \\
 K~\textsc{i}    & 19  & 7698.97   &   0.000  &   $-$0.170  &     47.6 \\
Ca~\textsc{i}    & 20  & 4435.68   &   1.890  &   $-$0.520  &     66.7 \\
\vdots           & \vdots & \vdots &  \vdots  &   \vdots    & \vdots   \\
\enddata
\tablecomments{
The full table is available in machine-readable form in the electronic
edition of the journal;
only a small portion is shown here to present
its general form and content.}
\end{deluxetable}

\clearpage

\begin{deluxetable}{ccccc}
%\tabletypesize{\scriptsize}
%\rotate
\tablecaption{Comparison of Model Atmosphere Parameters
\label{atmcompare}}
\tablewidth{0pt}
\tablecolumns{5}
\tablehead{
\colhead{Reference}      &
\colhead{$T_{\rm eff}$}  &
\colhead{log($g$)}       &
\colhead{$v_{\rm micro}$} & 
\colhead{[Fe/H]}         \\
\colhead{}               &
\colhead{(K)}            &
\colhead{}               &
\colhead{(\kmsec)}       &
\colhead{}               
}
\startdata
\multicolumn{5}{c}{HD 122563} \\
\hline
This study         & 4570 & 0.85 & 2.0  & $-$2.81 \\
\citet{aoki05}     & 4600 & 1.1  & 2.2  & $-$2.62 \\
\citet{barbuy03}, model 1 & 4600 & 1.5 & 2.0 & $-$2.71 \\
\citet{barbuy03}, model 2 & 4600 & 1.1 & 2.0 & $-$2.80 \\
\citet{fulbright00}& 4425 & 0.6  & 2.75 & $-$2.60 \\
\citet{fulbright03}& 4650 & 1.24 & 1.85 & $-$2.63 \\
\citet{honda04b}   & 4570 & 1.1  & 2.2  & $-$2.77 \\
\citet{johnson02a} & 4450 & 0.50 & 2.3  & $-$2.65 \\
\citet{mishenina01}& 4570 & 1.1  & 1.2  & $-$2.42 \\
\citet{roederer08} & 4570 & 0.55 & 2.4  & $-$2.83 \\
\citet{takeda-hidai05} & 4650 & 1.36 & 1.9  & $-$2.65 \\
\citet{thevenin98} & 4582 & 0.8  & 2.4  & $-$2.60 \\
\citet{westin00}   & 4500 & 1.3  & 2.5  & $-$2.70 \\
\hline
mean               & 4564 & 1.03 & 2.14 & $-$2.66 \\
standard dev.\     &   69 & 0.33 & 0.40 &    0.11 \\
\hline
\multicolumn{5}{c}{HD 84937} \\
\hline
This study         & 6300 & 4.0  & 1.2  & $-$2.28 \\
\citet{bihain04}   & 6277 & 4.03 & 1.0  & $-$2.06 \\
\citet{fulbright00}& 6375 & 4.1  & 0.8  & $-$2.0  \\
\citet{gratton03}  & 6290 & 4.02 & 1.25 & $-$2.18 \\
\citet{jonsell05}  & 6310 & 4.04 & 1.5  & $-$1.96 \\
\citet{mishenina01}& 6250 & 3.8  & 1.5  & $-$2.00 \\
\citet{nissen07}   & 6357 & 4.07 & 1.5  & $-$2.11 \\
\citet{smith93}    & 6090 & 4.0  & 1.5  & $-$2.4  \\
\citet{thevenin98} & 6222 & 4.0  & 1.3  & $-$2.10 \\
\citet{zhang05}    & 6261 & 4.07 & 1.8  & $-$1.93 \\
\hline
mean               & 6270 & 4.01 & 1.35 & $-$2.08 \\
standard dev.\     &   75 & 0.09 & 0.30 &    0.14 \\
\enddata
\tablecomments{The mean and standard deviation calculations
do not include the values derived in this study.}
\end{deluxetable}

\clearpage

\begin{deluxetable}{cccccccccc}
\tablecaption{Abundance Summary for \ligiant\
\label{abund}}
\tablewidth{0pt}
\tablecolumns{8}
\tablehead{
\colhead{} &
\colhead{LTE} &
\colhead{NLTE} &
\colhead{LTE} &
\colhead{} &
\colhead{} &
\colhead{} &
\colhead{}   \\
\colhead{Species} &
\colhead{log $\varepsilon$} &
\colhead{log $\varepsilon$} &
\colhead{[X/Fe]\tablenotemark{a}} &
\colhead{$\sigma$} &
\colhead{No.\ lines} &
\colhead{log $\varepsilon_{\sun}$\tablenotemark{a}} &
\colhead{Notes} 
}
\startdata
Fe \textsc{i}     &    5.27 & \nodata   & $-$2.23\tablenotemark{b} & 0.23 & 86 & 7.50 & EW    \\
Fe \textsc{ii}    &    5.27 & \nodata   & $-$2.23\tablenotemark{b} & 0.17 &  6 & 7.50 & EW    \\
Li \textsc{i}     &    2.06 &      2.16 & \nodata                  & 0.16 &  1 & 1.10 & synth \\
C                 &    6.97 & \nodata   & $+$0.68                  & 0.3  & \nodata & 8.52 & synth \\
O  \textsc{i}     &    7.56 & $\sim$7.2 & $+$1.06                  & 0.27 &  3 & 8.73 & EW    \\
Na \textsc{i}     &    4.79 & $\sim$4.7 & $+$0.69                  & 0.16 &  2 & 6.33 & EW    \\
Mg \textsc{i}     &    5.77 & $\sim$5.9 & $+$0.42                  & 0.25 &  4 & 7.58 & EW    \\
K  \textsc{i}     &    3.33 & $\sim$2.9 & $+$0.44                  & 0.17 &  1 & 5.12 & EW    \\
Ca \textsc{i}     &    4.56 & \nodata   & $+$0.43                  & 0.19 & 13 & 6.36 & EW    \\
Sc \textsc{ii}    &    1.14 & \nodata   & $+$0.20                  & 0.21 &  3 & 3.17 & EW    \\
Ti \textsc{i}     &    3.01 & \nodata   & $+$0.22                  & 0.22 &  9 & 5.02 & EW    \\
Ti \textsc{ii}    &    3.12 & \nodata   & $+$0.33                  & 0.20 &  9 & 5.02 & EW    \\
Cr \textsc{i}     &    3.29 & \nodata   & $-$0.15                  & 0.16 &  4 & 5.67 & EW    \\
Mn \textsc{i}     &    2.85 & \nodata   & $-$0.31                  & 0.24 &  1 & 5.39 & EW    \\
Ni \textsc{i}     &    4.13 & \nodata   & $+$0.11                  & 0.18 &  3 & 6.25 & EW    \\
Sr \textsc{ii}    &    1.05 & \nodata   & $+$0.36                  & 0.26 &  1 & 2.92 & synth \\
Y  \textsc{ii}    &    0.18 & \nodata   & $+$0.17                  & 0.19 &  4 & 2.24 & synth \\
Zr \textsc{ii}    &    1.14 & \nodata   & $+$0.77                  & 0.27 &  2 & 2.60 & synth \\
Ba \textsc{ii}    &    0.76 & \nodata   & $+$0.86                  & 0.29 &  4 & 2.13 & synth \\
La \textsc{ii}    & $-$0.28 & \nodata   & $+$0.78                  & 0.21 &  6 & 1.17 & synth \\
Ce \textsc{ii}    &    0.52 & \nodata   & $+$1.17                  & 0.20 &  4 & 1.58 & synth \\
Pr \textsc{ii}    & $-$0.42 & \nodata   & $+$1.10                  & 0.20 &  3 & 0.71 & synth \\
Nd \textsc{ii}    &    0.31 & \nodata   & $+$1.04                  & 0.19 & 12 & 1.50 & synth \\
Sm \textsc{ii}    & $-$0.49 & \nodata   & $+$0.73                  & 0.22 &  3 & 1.01 & synth \\
Eu \textsc{ii}    & $-$1.24 & \nodata   & $+$0.48                  & 0.20 &  2 & 0.51 & synth \\
\enddata
%\tablecomments{}
\tablenotetext{a}{Solar photospheric values from Grevesse \& Sauval (2002)}
\tablenotetext{b}{[Fe/H] is shown here.}
\end{deluxetable}

\clearpage

\begin{deluxetable}{cccc}
\tablecaption{Radial Velocity Variations among Stars Exhibiting
$r+s$ Enrichment
\label{rvrs}}
\tablewidth{0pt}
\tablehead{
\colhead{star name} &
\colhead{radial velocity} &
\colhead{period} &
\colhead{references} \\
\colhead{} &
\colhead{variations?\tablenotemark{a}} &
\colhead{(days)} &
\colhead{}}
\startdata
\ligiant\           & N & \nodata & 1          \\
CS~22183$-$015      & Y & \nodata & 7, 9       \\
CS~22898$-$027      & N & \nodata & 3, 13      \\
CS~22948$-$027      & Y & 426.5   & 5, 13      \\
CS~22964$-$161      & Y & 252     & 16         \\
CS~29497$-$030      & Y & 342     & 12, 14, 15 \\
CS~29497$-$034      & Y & 4130    & 5          \\
CS~29526$-$110      & Y & \nodata & 3          \\
CS~30322$-$023      & N\tablenotemark{b} & \nodata & \nodata \\
CS~31062$-$012      & N & \nodata & 11         \\
CS~31062$-$050      & Y & \nodata & 3, 4       \\
HE~0024$-$2523      & Y & 3.41    & 10         \\
HE~0131$-$3953      & ? & \nodata & \nodata    \\
HE~0338$-$3945      & N & \nodata & 8          \\
HE~1046$-$1352      & ? & \nodata & \nodata    \\
HE~1105$+$0027      & ? & \nodata & \nodata    \\
HE~1405$-$0822      & ? & \nodata & \nodata    \\
HE~2148$-$1247      & Y & (long)  & 6          \\
LP~625$-$44         & Y & $\gtrsim 4400$ & 2   \\
\enddata
\tablenotetext{a}{
In the second column, a ``Y'' indicates that the
star exhibits radial velocity variations, an ``N''
indicates that several measurements over a period of time
have confirmed that the star does not exhibit radial velocity
variations, and a ``?'' indicates that fewer than
two radial velocity measurements have been reported in the 
literature.}
\tablenotetext{b}{
CS~30322-023 is in the TP-AGB phase and may exhibit a pulsation
period of 192 days with an amplitude of $\approx 3$\,km\,s$^{-1}$.
This may mask the detection of a long, low-amplitude orbital period,
which cannot be ruled out by the available observations
\citep{masseron06}.}
\tablerefs{
(1)~this study;
(2)~\citet{aoki00};
(3)~\citet{aoki02};
(4)~\citet{aoki03};
(5)~\citet{barbuy05};
(6)~\citet{cohen03};
(7)~\citet{cohen06};
(8)~\citet{jonsell06};
(9)~\citet{lai04};
(10)~\citet{lucatello03};
(11)~\citet{norris97};
(12)~\citet{preston00};
(13)~\citet{preston01};
(14)~\citet{sivarani04};
(15)~\citet{sneden03};
(16)~\citet{thompson08}}
\end{deluxetable}

\clearpage

\begin{deluxetable}{cccc}
\tablecaption{Enhanced Na Abundances among Stars Exhibiting $r+s$ Enrichment
\label{naabund}}
\tablewidth{0pt}
\tablehead{
\colhead{star name} &
\colhead{[Na/Fe]$_{\rm LTE}$} &
\colhead{[Na/Fe]$_{\rm NLTE}$} &
\colhead{reference} }
\startdata
\ligiant\           & $+$0.69  & $+$0.6  & 1 \\
CS~22898$-$027      & $+$0.17  & \nodata & 6 \\
CS~22948$-$027      & $+$0.57  & $+$0.07 & 3 \\
CS~29497$-$030      & $+$0.58  & \nodata & 4 \\
CS~29497$-$034      & $+$1.18  & $+$0.68 & 3 \\
CS~30322$-$023      & $+$1.29  & \nodata & 5 \\
CS~31062$-$012\tablenotemark{a} & $+$1.3 & $+$0.6 & 2 \\
\enddata
\tablenotetext{a}{CS~30322-023 = LP~706-7}
\tablerefs{
(1)~this study;
(2)~\citet{aoki07};
(3)~\citet{barbuy05};
(4)~\citet{ivans05};
(5)~\citet{masseron06};
(6)~\citet{preston01}}
\end{deluxetable}

\clearpage

\clearpage

\clearpage

\clearpage

\clearpage

\clearpage

\clearpage

\clearpage

\end{document}